\documentclass[twocolumn, 10pt, pre, aps, showpacs, reprint, amsmath, amssymb, superscriptaddress, nofootinbib]{revtex4-1}

\usepackage[pdftex]{graphicx}
\usepackage{subfigure}
\usepackage{wasysym}

\begin{document}

\newcommand{\kt}{k_\parallel}
\newcommand{\lzt}{q_z}
\newcommand{\atled}{\bm{\nabla}}
\newcommand{\dx}{\frac{\partial}{\partial_x}}
\newcommand{\dy}{\frac{\partial}{\partial_y}}
\newcommand{\dz}{\frac{\partial}{\partial_z}}
\newcommand{\dt}{\frac{\partial}{\partial_t}}
\newcommand{\sqrdt}{\frac{\partial^2}{\partial_t^2}}
\newcommand{\pbyp}[2]{\frac{\partial #1}{\partial #2}}
\newcommand{\dbyd}[2]{\frac{d #1}{d #2}}
\newcommand{\ex}{\vec{e}_x}
\newcommand{\ey}{\vec{e}_y}
\newcommand{\ez}{\vec{e}_z}
\newcommand{\besselj}[2]{\mathrm{J}_{#1}(#2)}
\newcommand{\besseljp}[2]{\mathrm{J'}_{#1}(#2)}
\newcommand{\besseljs}[1]{\mathrm{J}_{#1}}
\newcommand{\besseljsp}[1]{\mathrm{J}_{#1}'}
\newcommand{\besseljspp}[1]{\mathrm{J}_{#1}''}
\newcommand{\hankel}[3]{\mathrm{H}_{#1}^{(#2)}(#3)}
\newcommand{\hankelp}[3]{\mathrm{H'}_{#1}^{(#2)}(#3)}
\newcommand{\hankels}[2]{\mathrm{H}_{#1}^{(#2)}}
\newcommand{\hankelsp}[2]{\mathrm{H}_{#1}'^{(#2)}}
\newcommand{\hankelspp}[2]{\mathrm{H}_{#1}''^{(#2)}}
\newcommand{\laplace}{\Delta}
\newcommand{\neff}{n_{\mathrm{eff}}}
\newcommand{\fexp}{f_{\mathrm{expt}}}
\newcommand{\ftheo}{f_{\mathrm{calc}}}
\newcommand{\nexp}{\tilde{n}}
\newcommand{\Gtheo}{\Gamma_{\mathrm{calc}}}
\newcommand{\Gexp}{\Gamma_{\mathrm{expt}}}
\newcommand{\Grad}{\Gamma_{\mathrm{rad}}}
\newcommand{\Gabs}{\Gamma_{\mathrm{abs}}}
\newcommand{\Gant}{\Gamma_{\mathrm{ant}}}
\newcommand{\rhof}{\rho_{\mathrm{fluc}}}
\newcommand{\rhofscl}{\rhof^{\mathrm{scl}}}
\newcommand{\rhofSS}{\rhof^{(\mathrm{ss})}}
\newcommand{\rhofnr}[1]{\rho_{#1 n_r}}
\newcommand{\rhow}{\rho_{\mathrm{Weyl}}}
\newcommand{\Nweyl}{N_{\mathrm{Weyl}}}
\newcommand{\rhot}{\hat{\rho}}
\newcommand{\rhotscl}{\rhot_{\mathrm{scl}}}
\newcommand{\rhotSS}{\rhot^{(\mathrm{ss})}}
\newcommand{\rhotnr}[1]{\tilde{\rho}_{#1 n_r}}
\newcommand{\fmin}{f_{\mathrm{min}}}
\newcommand{\kmin}{k_{\mathrm{min}}}
\newcommand{\fmax}{f_{\mathrm{max}}}
\newcommand{\kmax}{k_{\mathrm{max}}}
\newcommand{\fcrit}{f_{\mathrm{crit}}}
\newcommand{\kcrit}{k_{\mathrm{crit}}}
\newcommand{\po}{\mathrm{po}}
\newcommand{\lpo}{\ell_\po}
\newcommand{\lpeak}{\ell_\mathrm{peak}}
\newcommand{\lpeakscl}{\lpeak^\mathrm{scl}}
\newcommand{\lmax}{\ell_{\mathrm{max}}}
\newcommand{\alphacrit}{\alpha_{\mathrm{crit}}}
\newcommand{\chico}{\chi_{\mathrm{co}}}
\newcommand{\reffig}[1]{\mbox{Fig.~\ref{#1}}}
\newcommand{\subreffig}[1]{\mbox{Fig. \subref{#1}}}
\newcommand{\refeq}[1]{\mbox{Eq.~(\ref{#1})}}
\newcommand{\refsec}[1]{\mbox{Sec.~\ref{#1}}}
\newcommand{\reftab}[1]{\mbox{Table \ref{#1}}}
\newcommand{\etal}{\textit{et al.\ }}
\newcommand{\dA}{A}
\newcommand{\dB}{B}
\newcommand{\FSR}{\mathrm{FSR}}
\newcommand{\dist}{D}
\newcommand{\depth}{l}
\renewcommand{\Re}[1]{\mathrm{Re}\left(#1\right)}
\renewcommand{\Im}[1]{\mathrm{Im}\left(#1\right)}

\hyphenation{re-so-nan-ce re-so-nan-ces ex-ci-ta-tion z-ex-ci-ta-tion di-elec-tric ap-pro-xi-ma-tion ra-dia-tion Me-cha-nics quan-tum pro-posed Con-cepts pro-duct Reh-feld ob-ser-va-ble Se-ve-ral rea-so-nable Ap-pa-rent-ly re-pe-ti-tions re-la-tive quan-tum su-per-con-duc-ting ap-pro-xi-mate cri-ti-cal}

\title{Application of a trace formula to the spectra of flat three-dimensional dielectric resonators}

\author{S. Bittner}
\affiliation{Institut f\"ur Kernphysik, Technische Universit\"at Darmstadt, D-64289 Darmstadt, Germany}
\author{E. Bogomolny}
\affiliation{Universit{\'e} Paris-Sud, CNRS, LPTMS, UMR8626, Orsay, F-91405, France}
\author{B. Dietz}
\email{dietz@ikp.tu-darmstadt.de}
\author{M. Miski-Oglu}
\affiliation{Institut f\"ur Kernphysik, Technische Universit\"at Darmstadt, D-64289 Darmstadt, Germany}
\author{A. Richter}
\email{richter@ikp.tu-darmstadt.de}
\affiliation{Institut f\"ur Kernphysik, Technische Universit\"at Darmstadt, D-64289 Darmstadt, Germany}
\affiliation{ECT*, Villa Tambosi, I-38123 Villazano (Trento), Italy}

\date{\today}

\begin{abstract}
The length spectra of flat three-dimensional dielectric resonators of circular shape were determined from a microwave experiment. They were compared to a semiclassical trace formula obtained within a two-dimensional model based on the effective index of refraction approximation and a good agreement was found. It was necessary to take into account the dispersion of the effective index of refraction for the two-dimensional approximation. Furthermore, small deviations between the experimental length spectrum and the trace formula prediction were attributed to the systematic error of the effective index of refraction approximation. In summary, the methods developed in this article enable the application of the trace formula for two-dimensional dielectric resonators also to realistic, flat three-dimensional dielectric microcavities and -lasers, allowing for the interpretation of their spectra in terms of classical periodic orbits. 
\end{abstract}


\maketitle

\section{\label{sec:intr}Introduction}
Open dielectric resonators have received great attention due to numerous applications \cite{Matsko2005}, e.g., as microlasers \cite{McCall1992, Chu1993, Kuwata-Gonokami1995, Lebental2007} or as sensors \cite{Fang2004, Armani2006, He2011}, and as paradigms of open wave-chaotic systems \cite{Hentschel2009}. The size of dielectric microcavities typically ranges from a few to several hundreds of wavelengths. Wave-dynamical systems that are large compared to the typical wavelength have been treated successfully with semiclassical methods. These provide approximate solutions in terms of properties of the corresponding classical system. In the case of dielectric cavities, the corresponding classical system is an open dielectric billiard. Inside the billiard rays travel freely while, when impinging the boundary, they are partially reflected and refracted to the outside according to Snell's law and the Fresnel formulas. The field distributions of resonance states of dielectric cavities can be localized on the periodic orbits (POs) of the corresponding billiard \cite{Gmachl2002, Harayama2003, Tureci2002a, Unterhinninghofen2008a} and the far-field characteristics of microlasers can be predicted from its ray dynamics \cite{Noeckel1994, Noeckel1997, Altmann2009}. Semiclassical corrections to the ray picture due to the Goos-H\"anchen shift \cite{Goos1947}, Fresnel filtering \cite{Tureci2002}, and curved boundaries \cite{Hentschel2002} are under investigation for a more precise understanding of the connections between ray and wave dynamics \cite{Rex2002, Altmann2008c, Unterhinninghofen2010a}. \newline
One of the most important tools in semiclassical physics are trace formulas, which relate the density of states of a quantum or wave-dynamical system to the POs of the corresponding classical system \cite{Gutzwiller1970, Gutzwiller1971, Brack2003}. Recently, a trace formula for two-dimensional (2D) dielectric resonators was developed \cite{Bogomolny2008, Hales2011}. The trace formula was successfully tested for resonator shapes with regular classical dynamics in experiments with 2D dielectric microwave resonators \cite{Bittner2010} and with polymer microlasers of various shapes \cite{Lebental2007, Bogomolny2011}. However, typical microlasers like those used in Refs.\ \cite{McCall1992, Chu1993, Kuwata-Gonokami1995, Lebental2007} are three-dimensional (3D) systems. While trace formulas for closed 3D electromagnetic resonators have been derived \cite{Balian1977, Frank1996} and tested \cite{Dembowski2002}, hitherto there is practically no investigation of the trace formula for 3D dielectric resonators. The main reason is the difficulty of the numerical solution of the full 3D Maxwell equations for real dielectric cavities. The case of flat microlasers is special since their in-plane extensions are large compared to the typical wavelength, whereas their height is smaller than or of the order of the wavelength. Even in this case complete numerical solutions are rarely performed. In practice, flat dielectric cavities are treated as 2D systems by introducing a so-called effective index of refraction \cite{Smotrova2005, Lebental2007} (see below). This approximation has been used in Refs.\ \cite{Lebental2007, Bogomolny2011} and a good overall agreement between the experiments and the theory was found. However, it is known \cite{Bittner2009} that this 2D approximation (called the $\neff$ model in the following) introduces certain uncontrolled errors. Even the separation between transverse electric and transverse magnetic polarizations intrinsic in this approach is not, strictly speaking, valid for 3D cavities \cite{Schwefel2005}. To the best of the authors' knowledge, no \textit{a priori} estimates of such errors are known even when the cavity height is much smaller than the wavelength. The purpose of the present work is the careful comparison of the experimental length spectra and the trace formula computed within the 2D $\neff$ approximation. Furthermore, the effect of the dispersion of the effective index of refraction on the trace formula is investigated as well as the need for higher-order corrections of the trace formula due to, e.g., curvature effects. The experiments were performed with two dielectric microwave resonators of circular shape and different thickness like in Ref.\ \cite{Bittner2009}. These are known to be ideal testbeds for the investigation of wave-dynamical chaos \cite{Richter1999, StoeckmannBuch2000} and have been used, e.g., in Refs.\ \cite{Schaefer2006, Bittner2009, Bittner2010, Unterhinninghofen2011}. The results of these microwave experiments can be directly applied to dielectric microcavities in the optical frequency regime if the ratio of the typical wavelength and the resonator extensions are similar. The paper is organized as follows. The experimental setup and the measured frequency spectrum are discussed in \refsec{sec:expSetup}. Section \ref{sec:theo} summarizes the $\neff$ model for flat 3D resonators, the semiclassical trace formula for 2D resonators and how these are combined here. The experimental length spectra are compared to this model in \refsec{sec:Lspekt} and \refsec{sec:conc} concludes with a discussion of the results.

\section{\label{sec:expSetup}Experimental setup for the measurement of frequency spectra}

\begin{figure}[!b]
\begin{center}
\subfigure[]{
	\includegraphics[width = 8 cm]{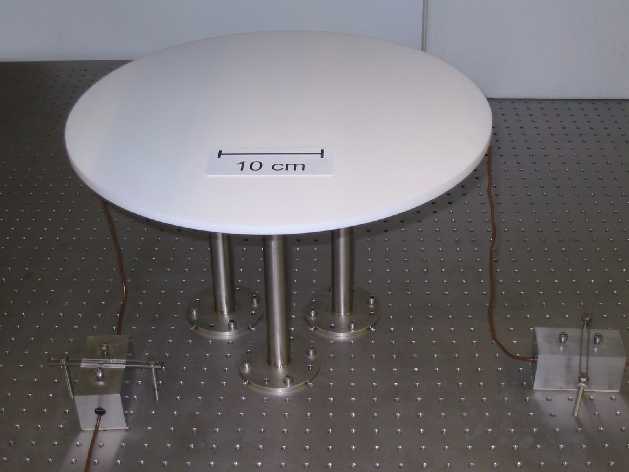}
	\label{sfig:setupPhoto}
}
\subfigure[]{
	\includegraphics[width = 8.4 cm]{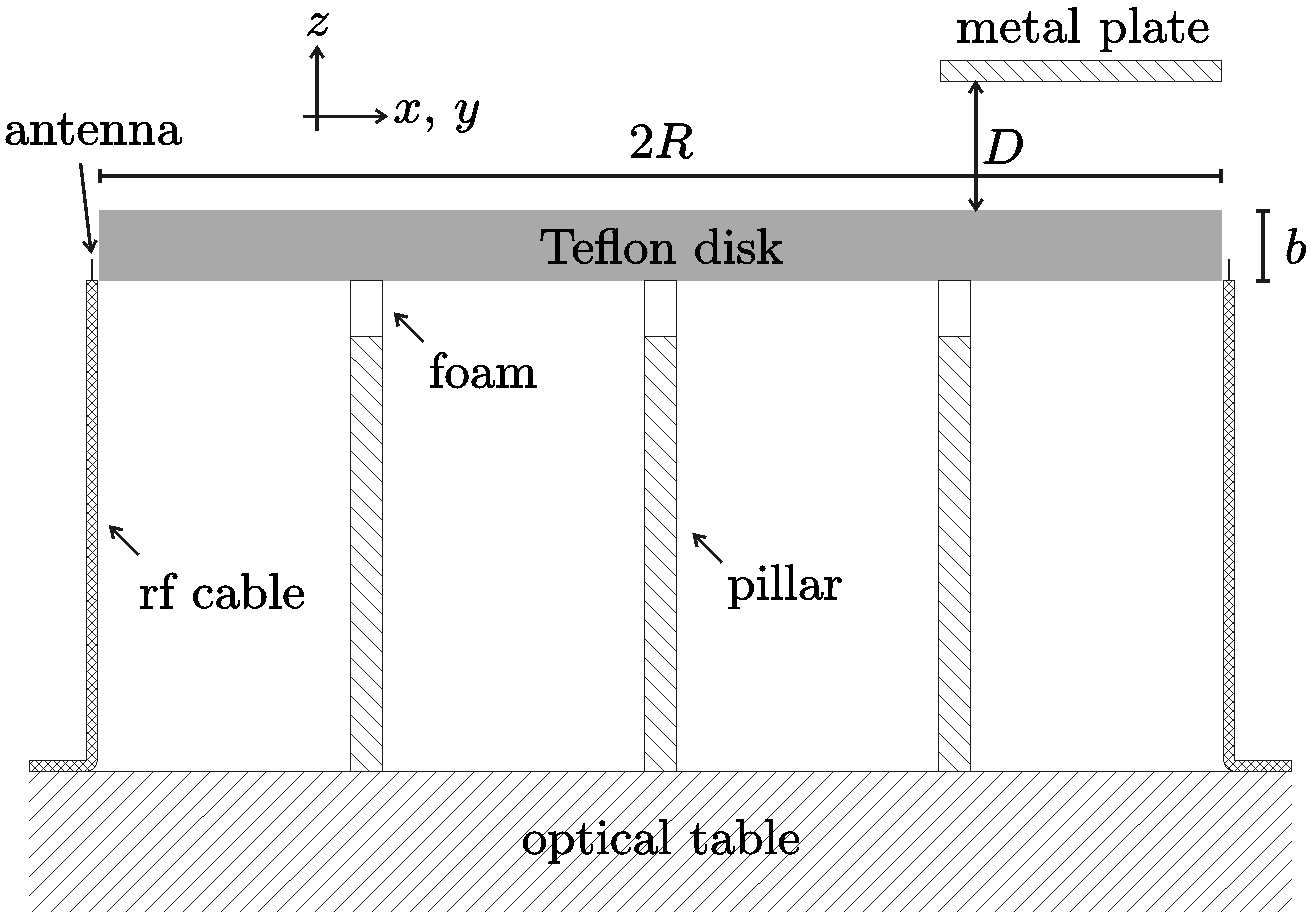}
	\label{sfig:setupSketch}
}
\end{center}
\caption{\label{fig:setup}(Color online) \subref{sfig:setupPhoto} Photograph of the experimental setup with disk \dB. \subref{sfig:setupSketch} Sketch of the setup (not to scale). Three metal pillars are used to support a Teflon disk with radius $R$ and thickness $b$. A special foam (see text) is used to isolate the disk from the pillars. Two vertical wire antennas protruding from coaxial RF cables are used to couple microwave power into and out of the disk. A metal plate with variable distance $\dist$ to the disk (not depicted in the photo) can be added to determine the polarization of the resonances.}
\end{figure}

Two flat circular disks made of Teflon were used as microwave resonators. The first one, disk \dA, has a radius of $R = 274.8$ mm and a thickness of $b = 16.7$ mm so $R / b = 16.5$. Its index of refraction is $n = 1.434$. A typical frequency of $f = 10$ GHz corresponds to $kR \approx 58$, where $k = 2 \pi f / c$ is the wave number and $c$ the speed of light. The second one, disk \dB, has $R = 225.1$ mm, $b = 10.1$ mm ($R / b = 22.3$), and $n = 1.430$, with $10$ GHz corresponding to $kR \approx 47$. The values of the indices of refraction $n$ of both disks were measured independently (see Ref.\ \cite{Bittner2009}) and validated by numerical calculations \cite{Classen2010}. They showed negligible dispersion in the considered frequency range \footnote{It was estimated that $\dbyd{n}{f} \times 10 \, \mathrm{GHz} \, < 1.5 \cdot 10^{-3}$.}. Figure \ref{sfig:setupPhoto} shows a photograph of the experimental setup. The circular Teflon disk is supported by three pillars arranged in a triangle. The prevalent modes observed experimentally are whispering gallery modes (WGMs) that are localized close to the boundary of the disk \cite{Bittner2009}. Therefore, the pillars perturb the resonator only negligibly because their position is far away from the boundary. Additionally, $4$ cm of a special foam (Rohacell 31IG by Evonik Industries \cite{Rohacell}) with an index of refraction of $n \approx 1.02$ is placed between the pillars and the disk as isolation [see \reffig{sfig:setupSketch}]. The total height of the pillars is $260$ mm so the resonator is not influenced by the optical table. Two vertical wire antennas are placed diametrically at the cylindrical sidewalls of the disk (cf.\ Ref.\ \cite{Bittner2009}). They are connected to a vectorial network analyzer (PNA N5230A by Agilent Technologies) with coaxial rf cables. The network analyzer measures the complex scattering matrix element $S_{21}(f)$, where
\begin{equation} |S_{21}(f)|^2 = \frac{P_{2, \mathrm{out}}}{P_{1, \mathrm{in}}} \end{equation}
is the ratio between the powers $P_{2, \mathrm{out}}$ coupled out via antenna $2$ and $P_{1, \mathrm{in}}$ coupled in via antenna $1$ for a given frequency $f$. Plotting $|S_{21}|^2$ versus $f$ yields the frequency spectrum.  

\begin{figure*}[tb]
\begin{center}
\includegraphics[width = 16 cm]{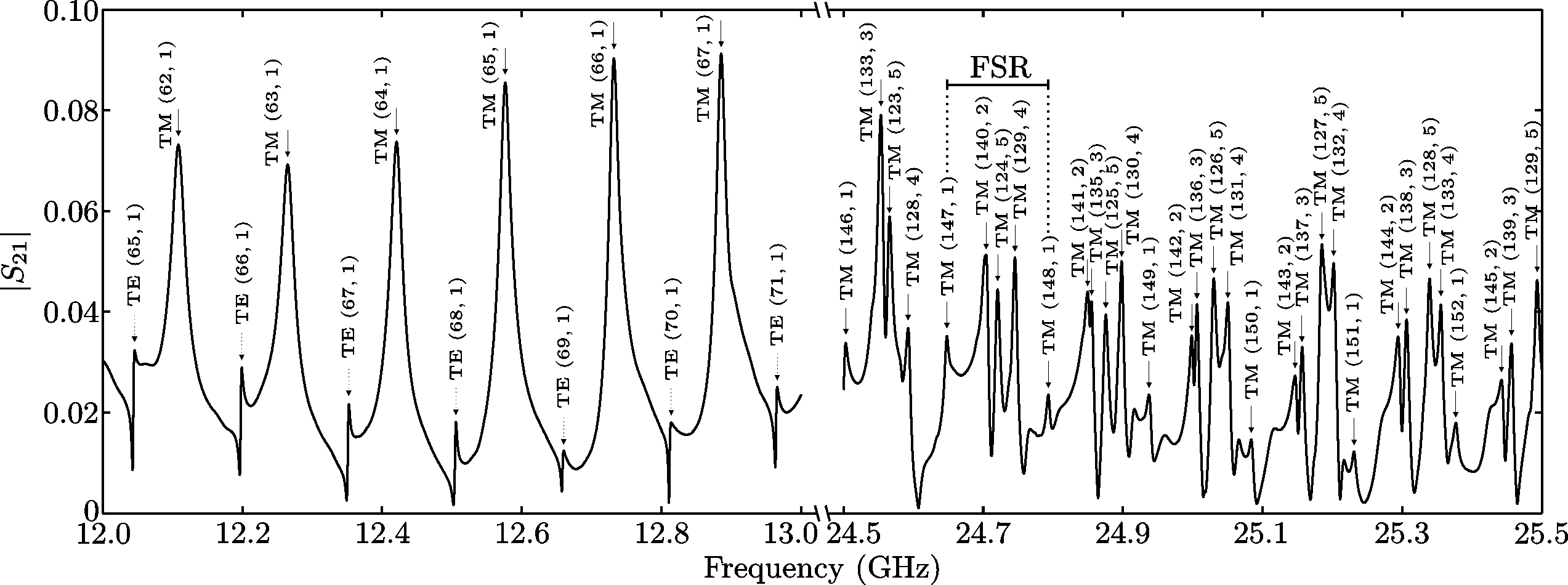}
\end{center}
\caption{\label{fig:fspec}Measured frequency spectrum of disk \dB. The polarization and quantum numbers are indicated by TM or TE $(m, n_r)$. In the right part, only the relevant TM modes are indicated. The free spectral range ($\FSR$) is the spacing between resonances with the same polarization and radial quantum number.}
\end{figure*}

\begin{figure}[tb]
\begin{center}
\includegraphics[width = 8.4 cm]{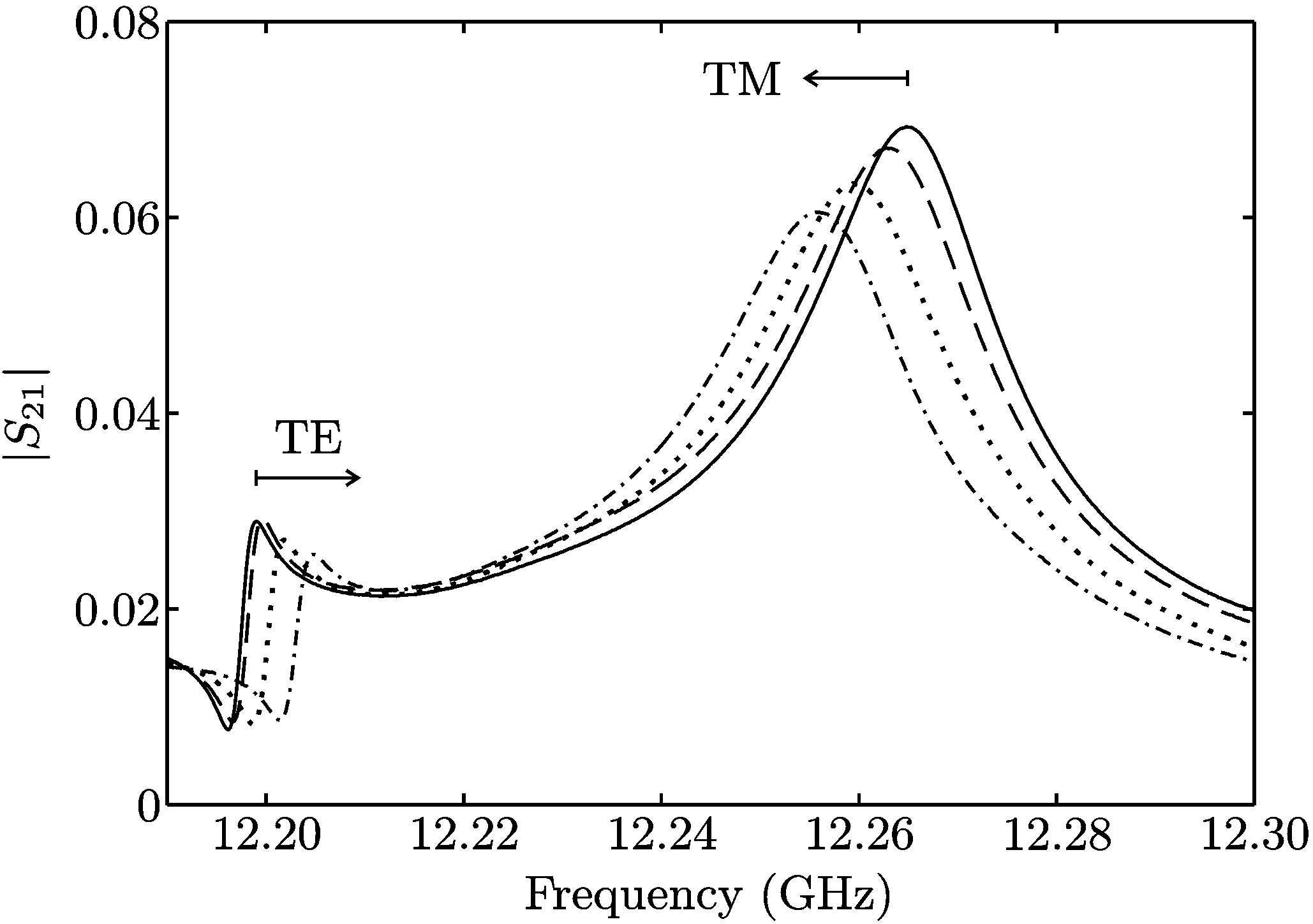}
\end{center}
\caption{\label{fig:polMeas}Frequency spectrum without metal plate (solid line) and with metal plate [see \reffig{sfig:setupSketch}] at distances $\dist = 14$ mm (dashed), $10$ mm (dotted), and $8$ mm (dot-dashed) for disk \dB. With decreasing distance $D$ the TE modes are shifted to higher frequencies, while the TM modes are shifted to lower frequencies.}
\end{figure}

The measured frequency spectrum of disk \dB~is shown in \reffig{fig:fspec}. It consists of several series of roughly equidistant resonances. The associated modes can be labeled with their polarization and the quantum numbers of the circle resonator, which are indicated in \reffig{fig:fspec}. Here, $(m, n_r)$ are the azimuthal and radial quantum number, respectively; TM denotes transverse magnetic polarization with the $z$ component of the magnetic field, $B_z$, equal to zero; and TE denotes transverse electric polarization with the $z$ component of the electric field, $E_z$, equal to zero. Each series of resonances consists of modes with the same polarization and radial quantum number. The free spectral range ($\FSR$) for each series is in the range of $145$--$170$ MHz. Only modes with $n_r \leq 5$, that is, WGMs, are observed in the experiment. The quantum numbers were determined from the intensity distributions, which were measured with the perturbation body method (see Ref.\ \cite{Bittner2009} and references therein). To determine the polarization of the modes a metal plate was placed parallel to the resonator at a variable distance $\dist$ [see \reffig{sfig:setupSketch}]. Figure~\ref{fig:polMeas} shows a part of the frequency spectrum with two resonances for different distances of the metal plate to the disk. The metal plate induces a shift of the resonance frequencies, where the magnitude of the frequency shift increases with decreasing distance $\dist$. Notably, the direction of the shift depends on the polarization of the corresponding mode: TE modes are shifted to higher frequencies and TM modes to lower ones, so the polarization of each mode can be determined uniquely. This behavior is attributed to the different boundary conditions for the $E_z$ field (TM modes) and the $B_z$ field (TE modes) at the metal plates. The former obeys Dirichlet, the latter Neumann boundary conditions. A detailed explanation is given in Appendix \ref{sec:polMeasCalc}.

\section{\label{sec:theo}The effective index of refraction and the trace formula}

\begin{figure}[bt]
\begin{center}
\includegraphics[width = 8 cm]{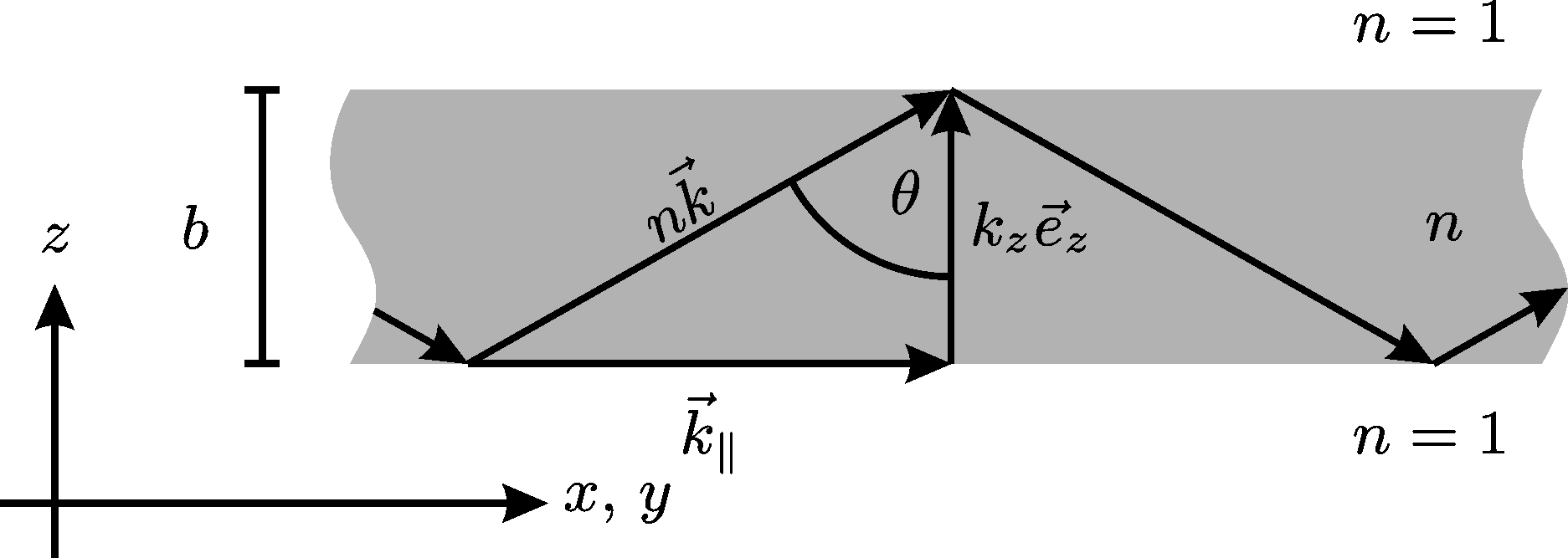}
\end{center}
\caption{\label{fig:rayGeom}Ray traveling through an infinite dielectric slab with thickness $b$ and index of refraction $n$. The wave vector $\vec{k}$ is decomposed into its components perpendicular ($k_z \ez$) and parallel ($\vec{k}_\parallel$) to the plane of the disk, where $k_\parallel = \neff k$. The angle of incidence on the top and bottom surface is $\theta$.}
\end{figure}

The open dielectric resonators are described by the vectorial Helmholtz equation
\begin{equation} \label{eq:helmholtzVec} [\Delta + n^2(\vec{r}) k^2] \left\{ \begin{array}{c} \vec{E} \\ \vec{B} \end{array} \right\} = \vec{0} \end{equation}
with outgoing-wave boundary conditions, where $\vec{E}$ and $\vec{B}$ are the electric and the magnetic field, respectively, and $n(\vec{r})$ is the index of refraction at the position $\vec{r}$. Though all components of the electric and magnetic fields obey the same Helmholtz equation, they are not independent but rather coupled in the bulk and at the boundaries as required by the Maxwell equations. The eigenvalues $k_j$ of \refeq{eq:helmholtzVec} are complex and the real part of $k_j$ corresponds to the resonance frequency $f_j = c \, \Re{k_j} / (2 \pi)$ of the resonance $j$, while the imaginary part corresponds to the resonance width $\Gamma_j = -c \, \Im{k_j} / \pi$ (full width at half maximum).  

\begin{figure}[tb]
\begin{center}
\includegraphics[width = 8.4 cm]{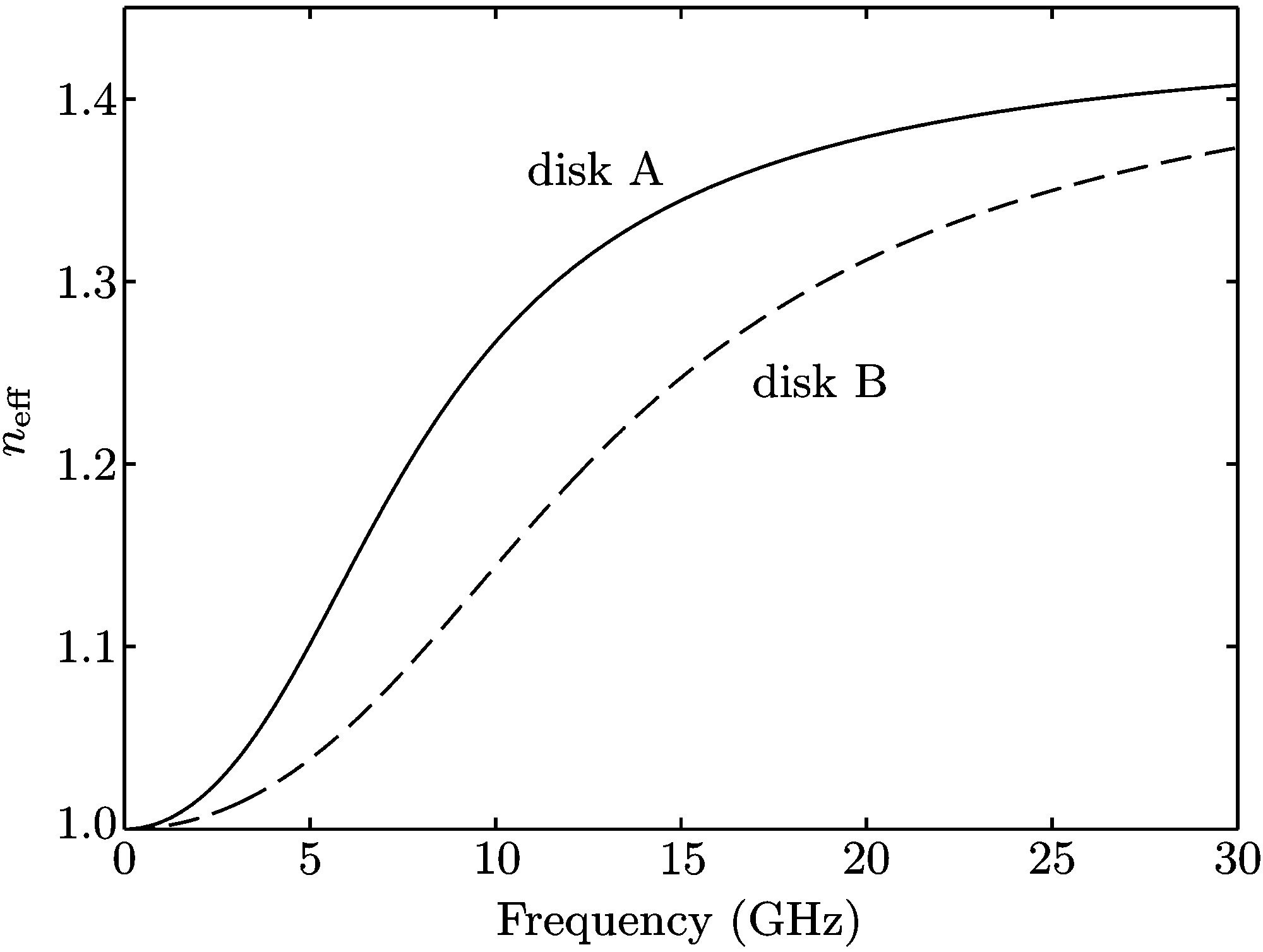}
\end{center}
\caption{\label{fig:neff}Effective index of refraction for the TM modes with $\zeta=0$ of disk \dA~(solid line) and disk \dB~(dashed line).}
\end{figure}

For the infinite slab geometry (see \reffig{fig:rayGeom}), the vectorial Helmholtz equation can be simplified by separating the wave vector $\vec{k}$ into a vertical $z$ component, $k_z \ez$, and a component parallel to the $x$-$y$ plane, $\vec{k}_\parallel$. Thus, $k^2 = (k_z^2 + \kt^2) / n^2$, and the angle of incidence on the top and bottom surface of the resonator is $\theta = \arctan{(\kt / k_z)}$. For a resonant wave inside the slab the wave vector component $k_z$ must obey the resonance condition
\begin{equation} \label{eq:kzQbed} \exp{(2 i k_z b)} r^2(\theta) = 1 \, , \end{equation}
where $r$ is the Fresnel reflection coefficient. The solutions of \refeq{eq:kzQbed} yield the quantized values of $k_z$. The effective index of refraction $\neff$ is defined as 
\begin{equation} \neff = n \sin{\theta} = \frac{\kt}{k}  \end{equation}
and corresponds to the phase velocity with respect to the $x$-$y$ plane. In the experiments only modes trapped due to total internal reflection (TIR) are observed. In this case, the reflection coefficient $r$ can be written as 
\begin{equation} \label{eq:rAsPhase} r = \exp{(-2 i \delta_0)} \end{equation}
with
\begin{equation} \label{eq:fresnelPhase} \tan{\delta_0} = \nu \sqrt{\frac{\neff^2 - 1}{n^2 - \neff^2}} \end{equation}
where $\nu = n^2$ for TM modes and $\nu = 1$ for TE modes. With these definitions, the quantization condition \refeq{eq:kzQbed} for $k_z$ can be reformulated as an implicit equation for the determination of $\neff$,
\begin{equation} \label{eq:neffQbed} kb = \frac{1}{\sqrt{n^2 - \neff^2}} \left[ 2 \arctan{\left( \nu \sqrt{\frac{\neff^2 - 1}{n^2 - \neff^2}} \right)} + \zeta \pi \right] \end{equation}
with $\zeta = 0, 1, 2, \dots$ being the order of excitation in the $z$ direction \cite{Lebental2007}. The $\arctan$ term in \refeq{eq:neffQbed} corresponds to the Fresnel phase due to the reflections and the $\zeta \pi$ term to the geometrical phase. In the framework of the $\neff$ model the flat resonator is treated as a dielectric slab waveguide and the vectorial Helmholtz equation [\refeq{eq:helmholtzVec}] is accordingly reduced to the 2D scalar Helmholtz equation \cite{Smotrova2005, Lebental2007, Bittner2009} by replacing $n$ by the effective index of refraction $\neff$,
\begin{equation} \label{eq:helmholtzScal} \begin{array}{clcl} ( \Delta + \neff^2 k^2 ) & \Psi_\mathrm{in}(x, y) & = & 0 \\ ( \Delta + k^2 ) & \Psi_\mathrm{out}(x, y) & = & 0 \, , \end{array} \end{equation}
where the wave function $\Psi_\mathrm{in, out}$ inside, respectively, outside of the resonator corresponds to $E_z$ in the case of TM modes and to $B_z$ in the case of TE modes. The boundary conditions at the boundary of the resonator in the $x$-$y$ plane (i.e., the cylindrical sidewalls), $\partial S$, are
\begin{equation} \Psi_\mathrm{in}|_{\partial S} = \Psi_\mathrm{out}|_{\partial S} \quad \mathrm{and} \quad \mu \left. \pbyp{\Psi_\mathrm{in}}{\vec{n}} \right|_{\partial S} = \left. \pbyp{\Psi_\mathrm{out}}{\vec{n}} \right|_{\partial S} \end{equation}
where $\vec{n}$ is the unit normal vector for $\partial S$, $\mu = 1$ for TM modes, and $\mu = 1 / \neff^2$ for TE modes. Equation (\ref{eq:helmholtzScal}) can be solved analytically for a circular dielectric resonator \cite{Hentschel2002b}. However, it should be stressed that \refeq{eq:helmholtzScal} is not exact for flat 3D cavities. It defines the 2D $\neff$ approximation whose accuracy is unknown analytically but which has been determined experimentally in Ref.\ \cite{Bittner2009}. Our purpose is to investigate the precision of this approximation for the length spectra of simple 3D dielectric cavities. The effective index of refraction for the TM modes with the lowest $z$ excitation $\zeta = 0$ of disk \dA~and \dB~is shown in \reffig{fig:neff}. Obviously, $\neff$ depends strongly on the frequency, and this dispersion plays a crucial role in the present work. It should be noted that also TE modes and modes with higher $z$ excitation exist in the considered frequency range, however, in the following we focus on TM modes.  

The density of states (DOS) in a dielectric resonator is given by \cite{Bogomolny2008}
\begin{equation} \label{eq:dos} \rho(k) = - \frac{1}{\pi} \sum \limits_j \frac{\Im{k_j}}{[k - \Re{k_j}]^2 + [\Im{k_j}]^2} \, , \end{equation}
where the summation runs over all resonances $j$. The DOS can be separated into a smooth, average part $\bar{\rho}$ and a fluctuating part, $\rhof$. The smooth part is well described by the Weyl formula given in Ref.\ \cite{Bogomolny2008} and depends only on the area, the circumference and the index of refraction of the resonator. The fluctuating part, on the other hand, is related to the POs of the corresponding classical dielectric billiard. For a 2D dielectric resonator with regular classical dynamics, the semiclassical approximation for $\rhof$ is \cite{Bogomolny2008}
\begin{equation} \label{eq:trFormScl} \rhofscl(k) = \sqrt{\frac{n^3}{\pi^3}} \sum \limits_\po B_\po |R_\po| \sqrt{k} \, e^{ i (n k \lpo + \varphi_\po) } + \mathrm{c.c.} \, \end{equation}
where $\lpo$ is the length of the PO, $R_\po$ is the product of the Fresnel reflection coefficients for the reflections of the rays at the dielectric interfaces, $\varphi_\po$ denotes the phase changes accumulated at the reflections [i.e., $\arg{(R_\po)}$] and at the conjugate points of the corresponding PO, and the amplitude $B_\po$ is proportional to $A_\po / \sqrt{\lpo}$, where $A_\po$ is the area of the billiard covered by the family of the PO. It should be noted that this semiclassical formula fails to accurately describe contributions of POs with angle of incidence close to the critical angle for TIR, $\alphacrit = \arcsin{(1 / n)}$, as concerns the amplitude. Consequently, higher-order corrections to the trace formula need to be developed for these cases \cite{Bogomolny2008}. We restrict the discussion to the experimentally observed TM modes and compare the results with the trace formula obtained in Ref.\ \cite{Bogomolny2008}. To select the TM modes with $\zeta = 0$ the polarization and the quantum numbers of all resonances had to be determined experimentally as described in \refsec{sec:expSetup}. The trace formula for 2D resonators is applied to the flat 3D resonators considered here by inserting the frequency-dependent effective index of refraction $\neff(k)$ instead of $n$ into \refeq{eq:trFormScl}. To test the accuracy of the resulting trace formula we computed the Fourier transform (FT) of $\rhof$. In Ref.\ \cite{Lebental2007} it was shown that it is essential to fully take into account the dispersion of $\neff$ in the FT for a meaningful comparison of the resulting length spectrum with the geometric lengths of the POs. Therefore, we define
\begin{equation} \label{eq:FTdef} \begin{array}{rcl} \rhot(\ell) & = & \int \limits_{\kmin}^{\kmax} dk \rhof(k) \exp{\{ -i k \neff(k) \ell \}} \\ & = & \sum \limits_j \exp{\{ -i k_j \neff(k_j) \ell \}} - \mathrm{FT}\{ \bar{\rho}(k) \} \, , \end{array} \end{equation}
where the quantity $\ell$ is a geometrical length  and $|\rhot(\ell)|$ is, thus, called the length spectrum. We will compare it to the FT as defined in \refeq{eq:FTdef} using $\neff(k)$ instead of $n$ of the trace formula, $\rhotscl(\ell)$. The resonance parameters $k_j$ are obtained by fitting Lorentzians to the measured frequency spectra, and $k_{\mathrm{min, \, max}} = 2 \pi f_{\mathrm{min, \, max}} / c$ correspond to the frequency range considered. Since in a circle resonator the resonance modes with $m > 0$ are doubly degenerate, the measured resonances are counted twice each. Note that even though only the most long-lived resonances (i.e., the WGMs) are observed experimentally, and these comprise only a fraction of all resonances, a comparison of the experimental length spectrum with the trace formula is meaningful \cite{Bittner2010}.

\section{\label{sec:Lspekt}Comparison of experimental length spectra and trace formula predictions}

\begin{figure}[b]
\begin{center}
\includegraphics[width = 8.4 cm]{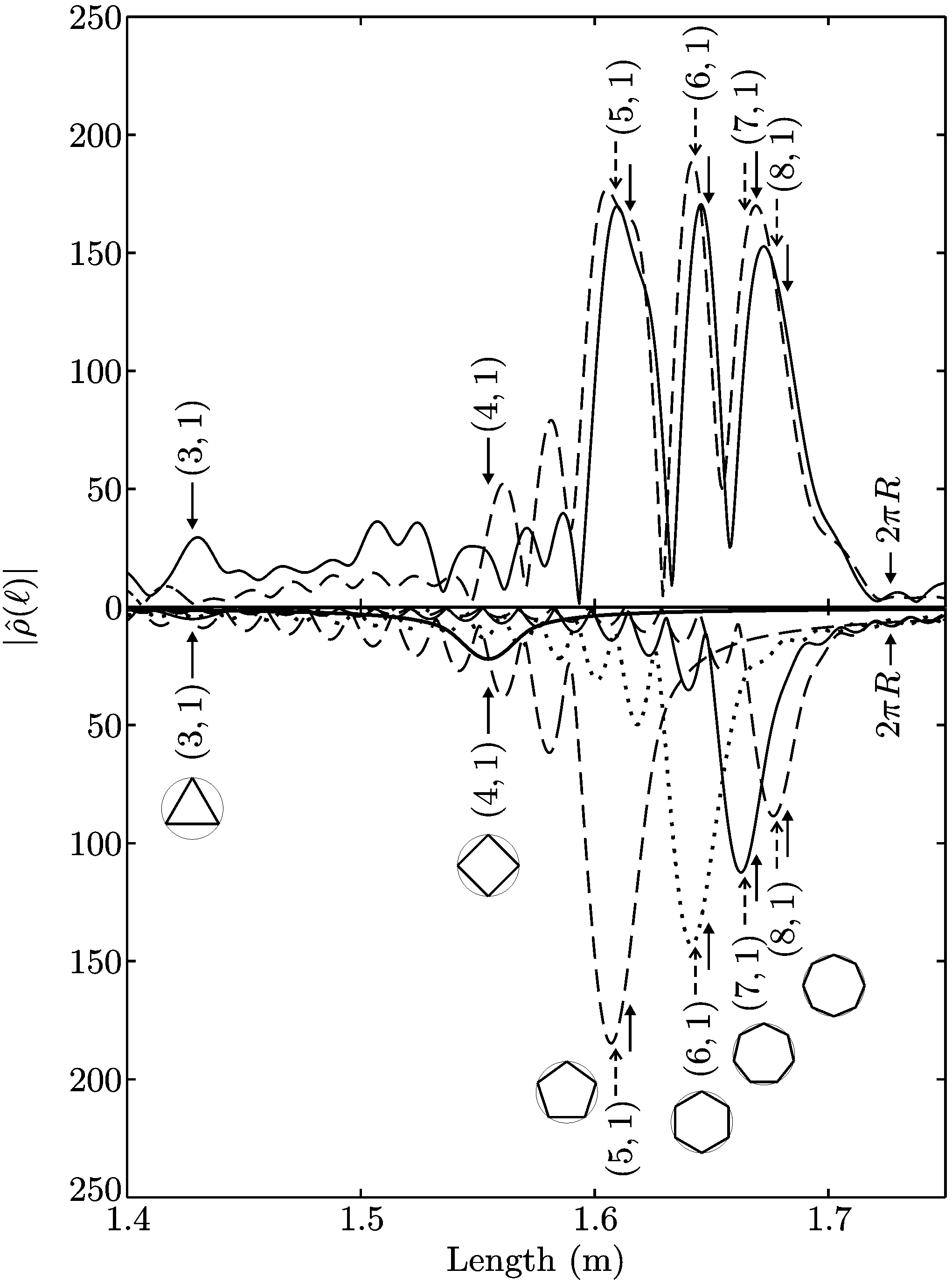}
\end{center}
\caption{\label{fig:lspektSclHKreis}Length spectrum for the TM modes of disk \dA. The solid line in the top part is the experimental length spectrum and the dashed line the FT of the semiclassical trace formula. The curves in the bottom part are the contributions of the individual POs to the semiclassical trace formula. The solid arrows indicate the lengths of the POs denoted by $(q, \eta)$ and the dashed arrows indicate the peak positions computed from \refeq{eq:peakPosEst}. The circumference $2 \pi R$ of the circle is also indicated.}
\end{figure}

\begin{figure}[tb]
\begin{center}
\includegraphics[width = 8.4 cm]{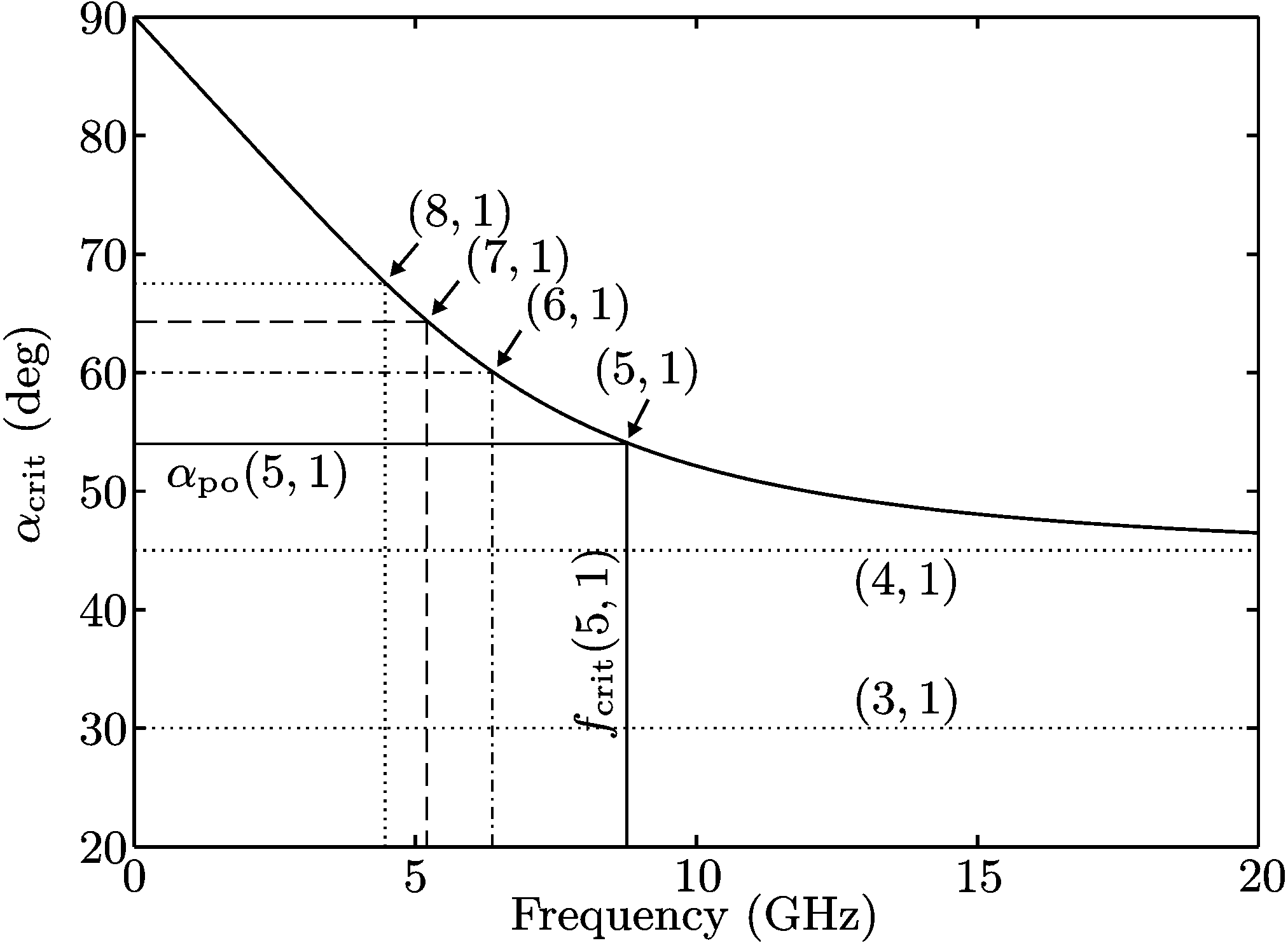}
\end{center}
\caption{\label{fig:aCritHKreis}Critical angle $\alphacrit = \arcsin{(1 / \neff)}$ for TIR with respect to the frequency for the TM modes of disk \dA. The angles of incidence of the $(q, \eta)$ orbits are indicated by the horizontal lines, and the critical frequencies $\fcrit$ at which they become confined by TIR by the vertical lines. The $(3, 1)$ and $(4, 1)$ orbits are not confined for the frequency range considered here.}
\end{figure}

\begin{figure}[tb]
\begin{center}
\includegraphics[width = 8.4 cm]{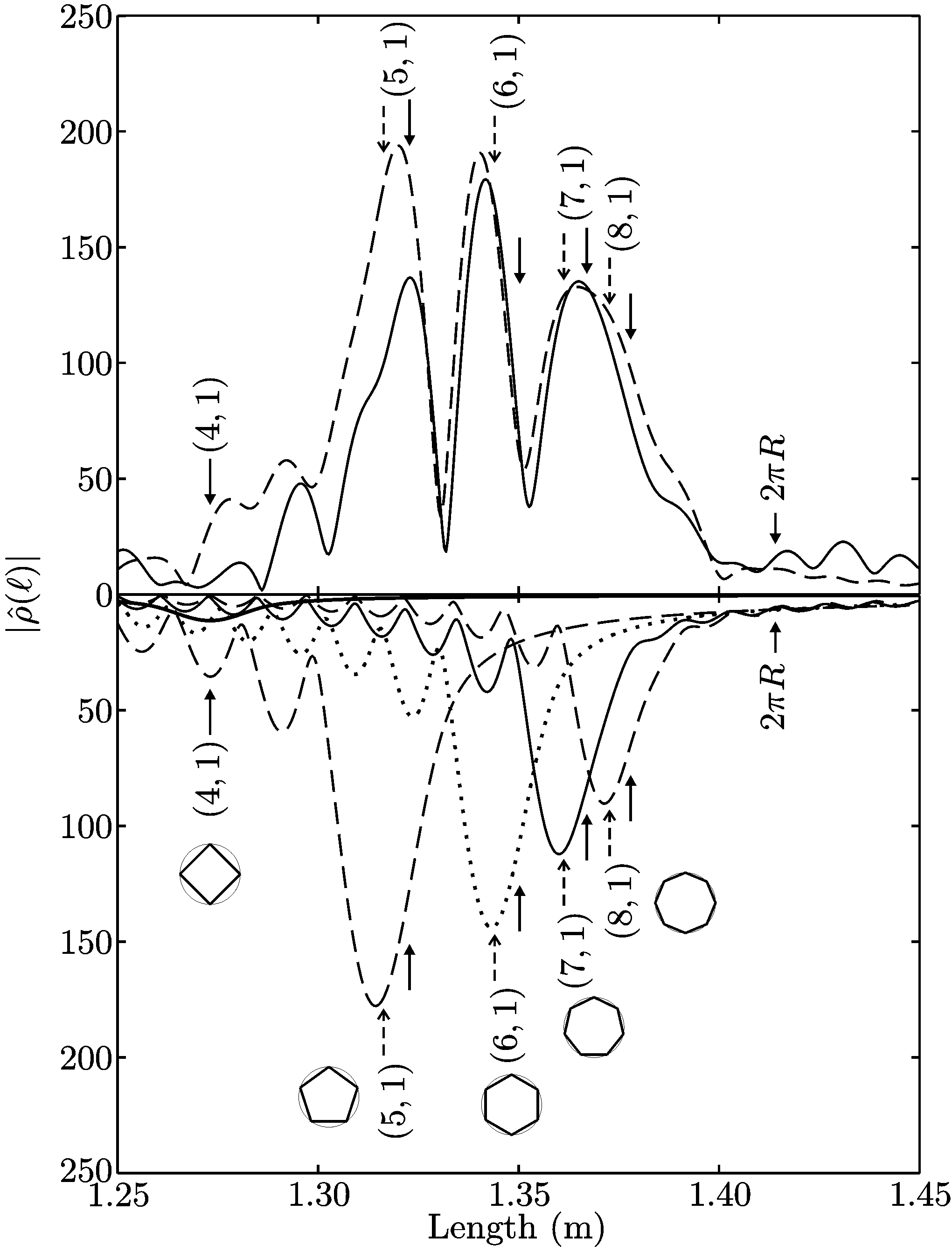}
\end{center}
\caption{\label{fig:lspektSclBKreis}Length spectrum for the TM modes of disk \dB. The solid line in the top part is the experimental length spectrum and the dashed line the FT of the semiclassical trace formula. The curves in the bottom part are the contributions of the individual POs to the semiclassical trace formula. The solid arrows indicate the lengths of the POs denoted by $(q, \eta)$ and the dashed arrows indicate the peak positions computed from \refeq{eq:peakPosEst}. The circumference $2 \pi R$ of the circle is also indicated.}
\end{figure}

Figure \ref{fig:lspektSclHKreis} shows the experimental length spectrum evaluated using \refeq{eq:FTdef} and the FT of the semiclassical trace formula, \refeq{eq:trFormScl}, for disk \dA. A total of $572$ measured TM modes with radial quantum numbers $n_r = 1$--$5$ from $\fmin = 6.8$ GHz to $\fmax = 20.0$ GHz was used. The POs in the circle billiard are denoted by their period $q$ and their rotation number $\eta$ and have the shape of polygons and stars (see insets in \reffig{fig:lspektSclHKreis}). Their lengths are indicated by the solid arrows. The POs with $2 \leq q \leq 15$ and $\eta = 1$ were used to compute the trace formula. Only POs with $q \leq 8$ are indicated in \reffig{fig:lspektSclHKreis} for the sake of clarity. The POs with $q \geq 9$ only add small contributions to the right shoulder of the peak corresponding to the $(7, 1)$ and $(8, 1)$ orbits. The amplitudes are
\begin{equation} B_\po = \frac{A_\po}{\sqrt{\lpo}} \sqrt{2} f_\po \quad \mathrm{with} \, f_\po = \left\{ \begin{array}{ll} 1, & q = 2 \eta \\ 2, & \mathrm{otherwise} \end{array} \right. , \end{equation}
and the phases are
\begin{equation} \label{eq:POphase} \varphi_\po = \frac{\pi}{4} - q \frac{\pi}{2} + q \arg{[r(\alpha_\po)]} \end{equation}
with the angles of incidence of the POs (with respect to the surface normal) being 
\begin{equation} \alpha_\po = \pi/2 - \eta \pi / q \, . \end{equation}
The overall agreement between the experimental length spectrum and the semiclassical trace formula is good and the major peaks in the length spectrum are close to the lengths of the $(5, 1)$--$(8, 1)$ orbits. However, no clear peaks are observed at the lengths of the $(3, 1)$ and the $(4, 1)$ orbit in the experimental length spectrum. Note that in the experimental length spectrum only orbits that are confined by TIR are observed (cf.\ Ref.\ \cite{Bittner2010}). This is not the case for the $(3, 1)$ and the $(4, 1)$ orbits in the frequency range of interest, where their angle of incidence $\alpha_\po$ is smaller than the critical angle $\alphacrit = \arcsin{(1 / \neff)}$ as depicted in \reffig{fig:aCritHKreis}. The length spectrum of disk \dB~is shown in \reffig{fig:lspektSclBKreis}. Altogether $546$ resonances with $n_r = 1$--$5$ from $\fmin = 9.8$ GHz to $\fmax = 26.9$ GHz were used. The semiclassical trace formula was again computed for $2 \leq q \leq 15$ and $\eta = 1$. The agreement of the experimental length spectrum and the FT of the trace formula is good and comparable to that obtained for disk \dA. As in the case of disk \dA, the experimental length spectrum exhibits no peaks for the $(3, 1)$ orbit, whose length is not within the range depicted in \reffig{fig:lspektSclBKreis}, and for the $(4, 1)$ orbit since they are not confined by TIR in the considered frequency range. A closer inspection of Figs.~\ref{fig:lspektSclHKreis} and \ref{fig:lspektSclBKreis} shows two unexpected effects. First, the peak positions of the FT of the semiclassical trace formula deviate slightly from the lengths of the POs. This can be seen best in the bottom parts of Figs.~\ref{fig:lspektSclHKreis} and \ref{fig:lspektSclBKreis}, where the contributions of the individual POs to the trace formula are depicted. Second, there is a small but systematic difference between the peak positions of the experimental length spectrum and those of the FT of the trace formula. We will demonstrate that the first effect is related to the dispersion of $\neff$ and the second effect to the systematic error of the $\neff$ model.

\begin{table}[tb]
\caption{Comparison of the PO lengths $\lpo$, the estimates for the peak positions $\lpeak$ from \refeq{eq:peakPosEst}, and the peak positions $\lpeakscl$ of the single orbit contributions to the semiclassical trace formula (see bottom parts of Figs.~\ref{fig:lspektSclHKreis} and \ref{fig:lspektSclBKreis}). The $(9, 1)$ and the $(10, 1)$ orbit are not indicated in Figs.~\ref{fig:lspektSclHKreis} and \ref{fig:lspektSclBKreis} since they give only small contributions to the rightmost peak in each figure. The top part is for disk \dA, the bottom one for disk \dB.}
\label{tab:lpeak}
\begin{center}
\begin{tabular}{cccc}
\hline \hline
$(q, \eta)$ & $\lpo$ (m) & $\lpeak$ (m) & $\lpeakscl$ (m) \\
\hline
Disk \dA \\
$(5, 1)$ & $1.615$ & $1.609$ & $1.607$ \\
$(6, 1)$ & $1.649$ & $1.643$ & $1.641$ \\
$(7, 1)$ & $1.669$ & $1.664$ & $1.663$ \\
$(8, 1)$ & $1.683$ & $1.678$ & $1.676$ \\
$(9, 1)$ & $1.692$ & $1.687$ & $1.686$ \\
$(10, 1)$ & $1.698$ & $1.694$ & $1.693$ \\
\\
Disk \dB \\
$(5, 1)$ & $1.323$ & $1.316$ & $1.314$ \\
$(6, 1)$ & $1.350$ & $1.344$ & $1.343$ \\
$(7, 1)$ & $1.367$ & $1.361$ & $1.360$ \\
$(8, 1)$ & $1.378$ & $1.373$ & $1.372$ \\
$(9, 1)$ & $1.385$ & $1.380$ & $1.379$ \\
$(10, 1)$ & $1.391$ & $1.386$ & $1.385$ \\
\hline
\hline
\end{tabular}
\end{center}
\end{table}

The difference between the peak positions of the trace formula and the lengths of the POs can be understood by considering the exponential term in the FT of the semiclassical trace formula \refeq{eq:trFormScl}, which for a single PO is 
\begin{equation} \label{eq:FTintegrand} \exp{\{ i [\neff(k) k (\lpo - \ell) + \varphi_\po] \}} \end{equation}
with $\varphi_\po$ given by \refeq{eq:POphase}. The crucial point is that the phase $\varphi_\po$ is frequency dependent because it contains the phase of the Fresnel coefficients, which, in turn, depends on $\neff(k)$. The modulus of the FT will be largest for that length $\ell$ for which the exponent in \refeq{eq:FTintegrand} is stationary, i.e., its derivative with respect to $k$ vanishes. This leads to the following estimate $\lpeak$ for the peak position,
\begin{equation} \label{eq:peakPosEst} \lpeak = \lpo - 2 q \left. \frac{\pbyp{\delta}{\neff}(\alpha_\po) \pbyp{\neff}{k}}{\neff + \pbyp{\neff}{k} k} \right|_{k_0} \, , \end{equation}
where
\begin{equation} \delta(\alpha, \neff) = \arctan{\left\{ \frac{\sqrt{\neff^2 \sin^2{\alpha} - 1}}{\neff \cos{\alpha}} \right\}} \end{equation}
is related to the Fresnel coefficients via \refeq{eq:rAsPhase}. The wave number $k_0$ at which \refeq{eq:peakPosEst} is evaluated is the center of the relevant wave number/frequency interval, which is
\begin{equation} k_0 = \left\{ \begin{array}{ccl} (\kmin + \kmax) / 2 & : & \kcrit < \kmin \\ (\kcrit + \kmax) /2 & : & \kcrit \geq \kmin \end{array} \right. \end{equation}
with $\kcrit = 2 \pi \fcrit / c$. The frequency $\fcrit$ is defined by $\sin{\alpha_\po} = 1 / \neff(\fcrit)$, i.e., it corresponds to the minimum frequency at which the considered PO is confined by TIR (cf.\ \reffig{fig:aCritHKreis}). Below $\kcrit$ the Fresnel phase vanishes. The estimated peak positions $\lpeak$ are indicated by the dashed arrows in Figs.~\ref{fig:lspektSclHKreis} and \ref{fig:lspektSclBKreis} and agree well with the peak positions $\lpeakscl$ of the individual PO contributions in the bottom parts of the figures. A list of the lengths of the POs, the peak positions of the single PO contributions, and the estimates $\lpeak$ according to \refeq{eq:peakPosEst} for disks \dA~and \dB~is provided in \reftab{tab:lpeak}. In general, the estimate $\lpeak$ deviates only by $1$--$2$ mm from the actual peak position $\lpeakscl$ (about $1 \permil$ of $\lpeakscl$). The $(3, 1)$ and the $(4, 1)$ orbits are not confined by TIR. Therefore, for these orbits the Fresnel phase vanishes and accordingly $\lpeak = \lpo$. Furthermore, their contributions to the length spectrum are symmetric with respect to $\ell = \lpo$, while those of the other POs are asymmetric with an oscillating tail to the left (see bottom parts of Figs.~\ref{fig:lspektSclHKreis} and \ref{fig:lspektSclBKreis}). These tails are attributed to the frequency dependence of the Fresnel phase. They can lead to interference effects, as can be seen for example in \reffig{fig:lspektSclBKreis}. There, e.g., the peak positions of the semiclassical trace formula (dashed line in the top part) for the $(5, 1)$ and the $(6, 1)$ orbit deviate from the peak positions $\lpeakscl$ of the corresponding single orbit contributions (dashed and dotted lines in the bottom part) due to interferences between the contributions of a PO and the side lobes of those of  the other POs. In order to identify such interferences, it is generally instructive to compare the FT of the semiclassical trace formula with those of its single orbit contributions. It should be noted that the effect discussed in this paragraph also occurs for any 2D resonator made of a dispersive material.  

It was shown in the previous paragraph that the dispersion of $\neff$ plays an important role. Furthermore, the semiclassical trace formula is known to be imprecise for POs with angles of incidence close to the critical angle. This is especially crucial here since several POs are close to the critical angle in at least a part of the considered frequency regime (see \reffig{fig:aCritHKreis}). These deficiencies of the semiclassical trace formula indicate the necessity to implement modifications of it. To pursue this presumption we will compare the experimental length spectrum with the FT of the exact trace formula for the 2D dielectric circle resonator using a frequency-dependent index of refraction $n(k)$ in order to investigate the deviations between it and the FT of the semiclassical trace formula. The trace formula is called exact since it is derived directly from the quantization condition for the dielectric circle resonator and without semiclassical approximations. It is given by
\begin{equation} \label{eq:trFormExact} \rhof(k) = \sum \limits_{\po = (q, \eta)} \frac{4}{\pi^2 k} \int \limits_{- \infty}^\infty dm \, e^{2 \pi i \eta m} P_m (R_m E_m)^q + \mathrm{c.c.} \end{equation}
with the definitions
\begin{equation} \label{eq:treEm} E_m(x) = \frac{\hankels{m}{1}}{\hankels{m}{2}}(nx) \, , \end{equation}
\begin{equation} \label{eq:trePm} P_m(x) = \frac{(n^2 - 1) (1 + \frac{k}{n} \dbyd{n}{k})}{A_m(x) B_m(x) \hankel{m}{1}{nx} \hankel{m}{2}{nx}} \, , \end{equation}
\begin{equation} \label{eq:treRm} R_m(x) = - \frac{A_m(x)}{B_m(x)} \, \end{equation}
\begin{equation} \label{eq:treAm} A_m(x) = n \frac{\hankelsp{m}{1}}{\hankels{m}{1}}(nx) - \frac{\hankelsp{m}{1}}{\hankels{m}{1}}(x) \end{equation}
and
\begin{equation} \label{eq:treBm} B_m(x) = n \frac{\hankelsp{m}{2}}{\hankels{m}{2}}(nx) - \frac{\hankelsp{m}{1}}{\hankels{m}{1}}(x) \, . \end{equation}
Here, $x = kR$, $\hankel{m}{1, 2}{x}$ are the Hankel functions of the first and second kinds, respectively, and the prime denotes the derivative with respect to the argument. Equation (\ref{eq:trFormExact}) is essentially Eq.~(67) of Ref.\ \cite{Bogomolny2008} with an additional factor $(1 + \frac{k}{n}\dbyd{n}{k})$ in the term $P_m(x)$. A detailed derivation is given in Appendix \ref{sec:exactTraceForm}. In the semiclassical limit, the term $R_m^q$ in \refeq{eq:trFormExact} turns into the product of the Fresnel reflection coefficients, the term $e^{2 \pi i \eta m} E_m^q$ into the oscillating term $e^{i n k \lpo}$, and $P_m$ contributes to the amplitude $B_\po$. For POs close to the critical angle, i.e., when the stationary point of the integrand in \refeq{eq:trFormExact} is $m_0 \approx n k R \sin{\alphacrit}$, the term $R_m^q$ varies rapidly with $m$, whereas it is assumed to change slowly in the stationary phase approximation used to derive the semiclassical trace formula \cite{Bogomolny2008}. Therefore, including curvature corrections in the Fresnel coefficients does not suffice for an accurate calculation of the contributions of these POs to $\rhof$. Consequently, we compute the integral entering in \refeq{eq:trFormExact} numerically.  

\begin{figure}[tb]
\begin{center}
\includegraphics[width = 8.4 cm]{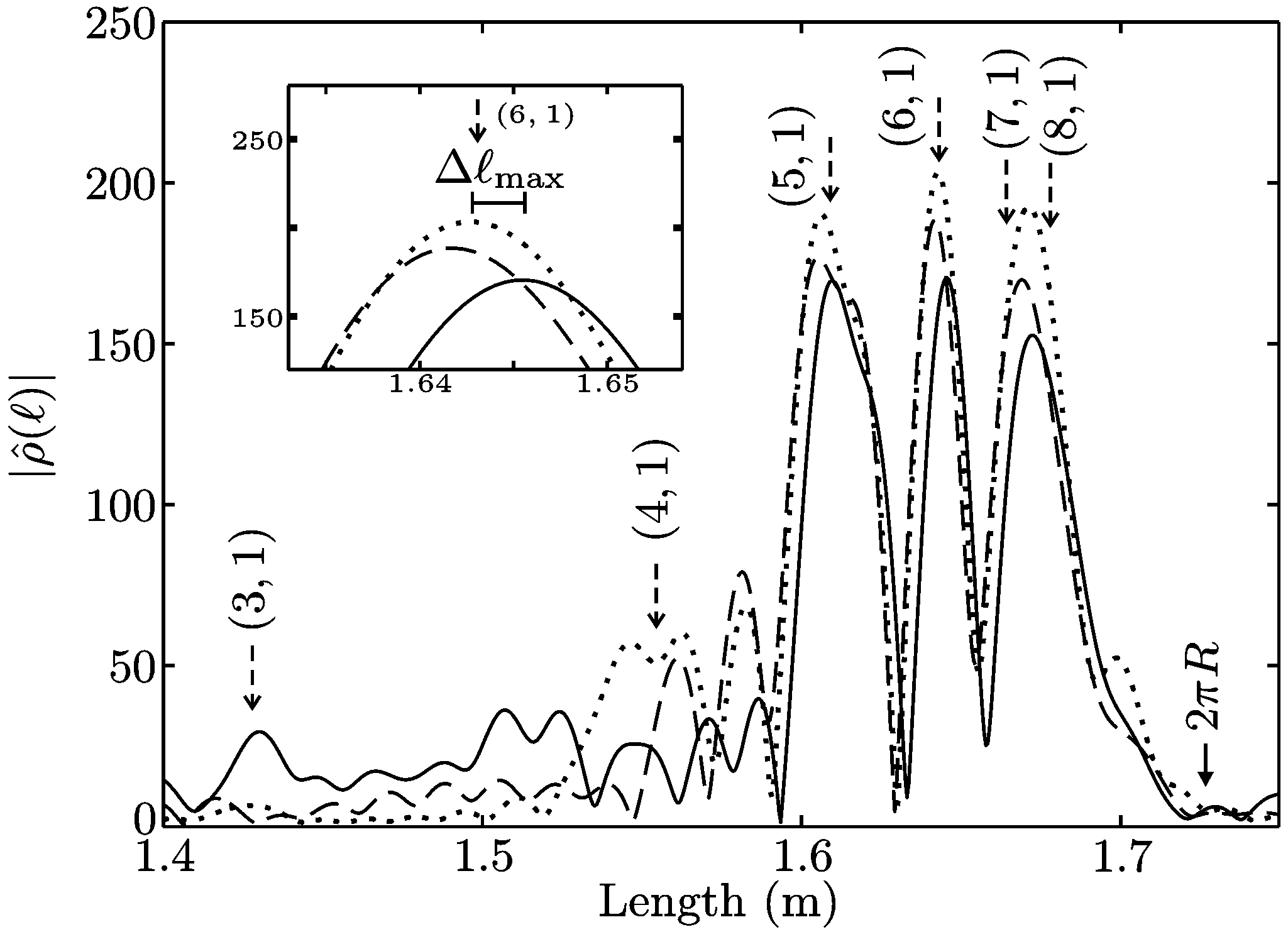}
\end{center}
\caption{\label{fig:lspektExactHKreis}Length spectrum for the TM modes of disk \dA. The solid line is the experimental length spectrum, the dashed line the FT of the semiclassical, and the dotted line the FT of the exact trace formula. The dashed arrows indicate the lengths $\lpeak$. A magnification around the peak corresponding to the $(6, 1)$ orbit is shown in the inset. Here, $\Delta \lmax$ denotes the difference between the positions of the maxima of the experimental length spectrum and those of the FT of the exact trace formula.}
\end{figure}

\begin{figure}[tb]
\begin{center}
\includegraphics[width = 8.4 cm]{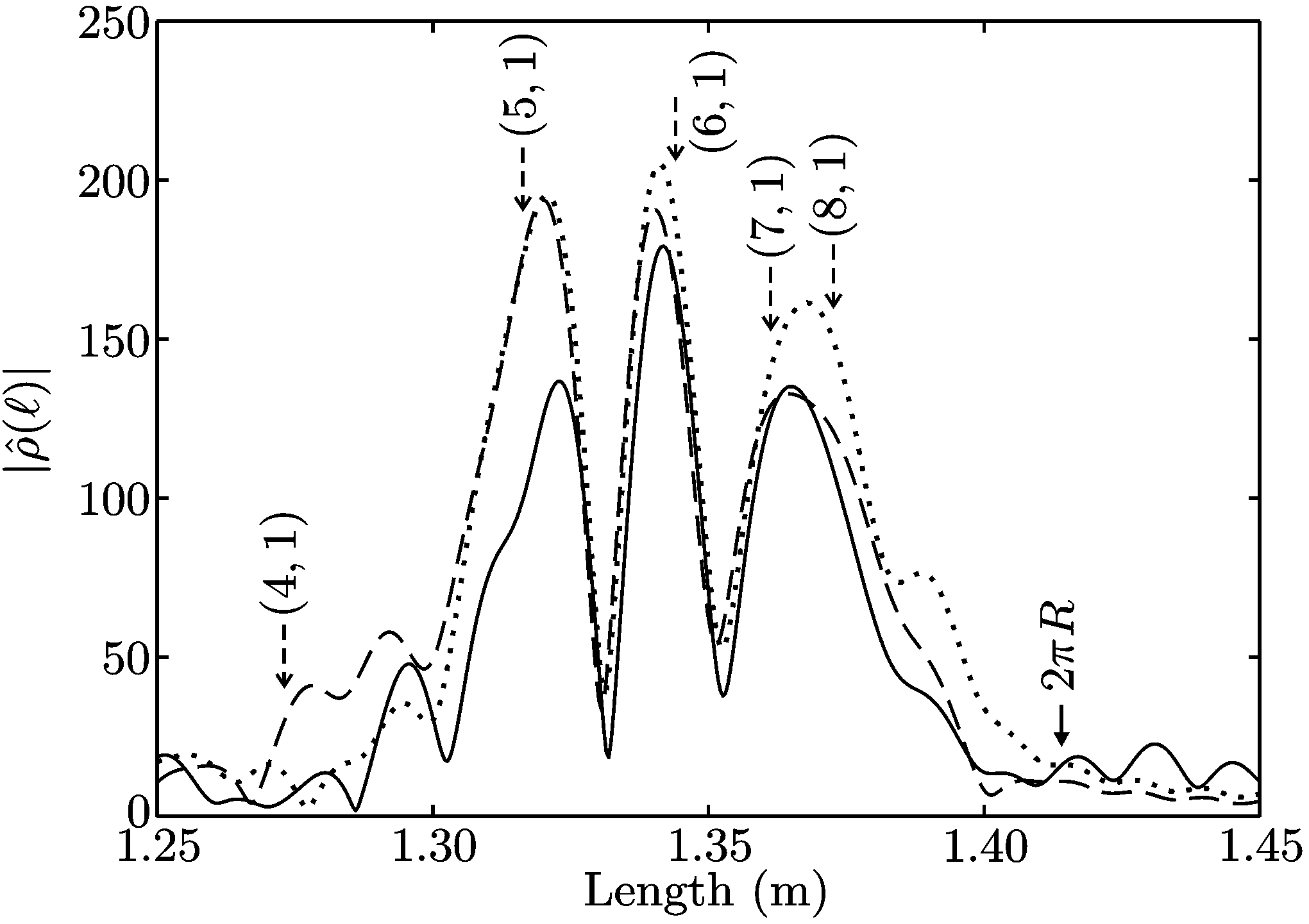}
\end{center}
\caption{\label{fig:lspektExactBKreis}Length spectrum for the TM modes of disk \dB. The solid line is the experimental length spectrum, the dashed line the FT of the semiclassical, and the dotted line the FT of the exact trace formula. The dashed arrows indicate the lengths $\lpeak$. The solid and dashed lines are the same as in the top part of \reffig{fig:lspektSclBKreis}.}
\end{figure}

\begin{figure}[bt]
\begin{center}
\includegraphics[width = 8.4 cm]{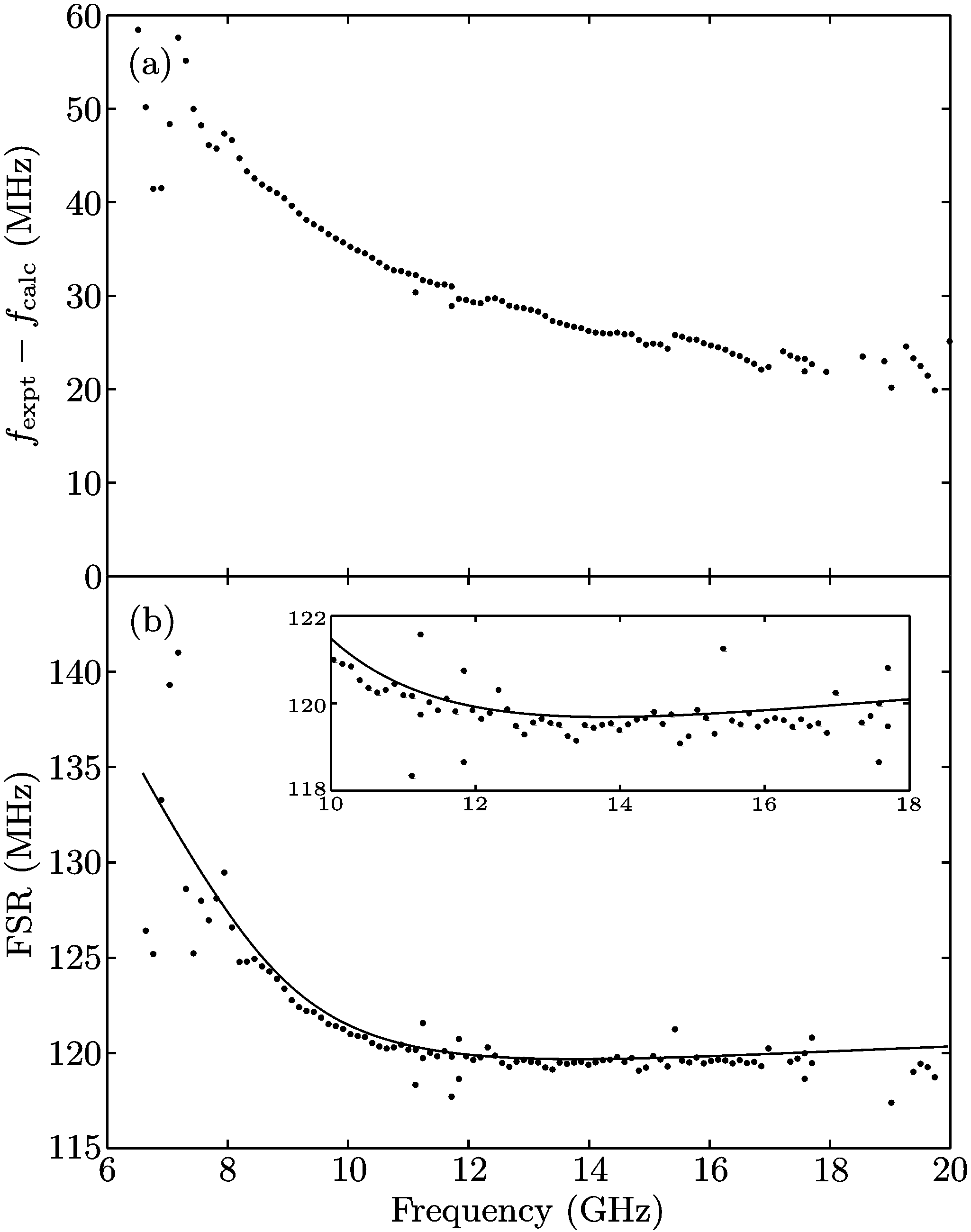}
\end{center}
\caption{\label{fig:freqFSRcomp}(a) Difference between the calculated and the measured resonance frequencies with respect to the measured frequency for the TM modes with $n_r = 1$ of disk \dA. The modes with $n_r > 1$ show a similar behavior (adapted from Ref.~\cite{Bittner2009}). (b) Measured (dots) and calculated (solid line) free spectral ranges for TM modes with $n_r = 1$. The calculated FSR is plotted as a curve instead of data points to guide the eye. The inset shows a magnification in the frequency range of $10$--$18$ GHz.}
\end{figure}

Figure \ref{fig:lspektExactHKreis} shows the comparison of the experimental length spectrum (solid line), the FT of the semiclassical trace formula (dashed line), \refeq{eq:trFormScl}, and that of the exact trace formula (dotted line), \refeq{eq:trFormExact}, for disk \dA. The latter was computed for $2 \leq q \leq 10$ and $\eta = 1$. The other two curves are the same as in the top part of \reffig{fig:lspektSclHKreis}. The main difference between the semiclassical and the exact trace formula are the larger peak amplitudes of the exact trace formula, where this is mainly attributed to the additional $(1 + \frac{k}{n} \dbyd{n}{k})$ factor. The differences between the peak amplitudes of the experimental length spectrum and those of the exact trace formula are actually expected since the measured resonances comprise only a part of the whole spectrum (cf. Ref.\ \cite{Bittner2010}), though they are not very large. On the other hand, the peak positions $\lmax^\mathrm{exact}$ of the exact trace formula differ only slightly from those of the semiclassical one, and still deviate from those of the experimental length spectrum, $\lmax^\mathrm{expt}$ (see inset of \reffig{fig:lspektExactHKreis}). The difference $\Delta \lmax = \lmax^\mathrm{expt} - \lmax^\mathrm{exact}$ is in the range of $3$--$4$ mm, i.e., about $2 \permil$ of the periodic orbit length $\lpo$. Similar effects are observed in \reffig{fig:lspektExactBKreis}, which shows the experimental length spectrum and the FT of the semiclassical trace formula and that of the exact trace formula (computed for $2 \leq q \leq 10$ and $\eta = 1$) for disk \dB. The peak amplitudes of the exact trace formula are, again, somewhat larger than those of the experimental length spectrum, and the peak positions $\lmax^\mathrm{exact}$ of the exact trace formula differ from those of the experimental length spectrum by $\Delta \lmax \approx 2$ mm for the $(5, 1)$ and the $(6, 1)$ orbits. The relative error $\Delta \lmax / \lmax \approx 1.5 \permil$ is, thus, slightly smaller than in the case of disk \dA.  

Since the trace formula itself is exact, the only explanation for these deviations is the systematic error of the $\neff$ model. Therefore, we compare the measured ($\fexp$) and the calculated ($\ftheo$) resonance frequencies for disk \dA~in \reffig{fig:freqFSRcomp}. The resonance frequencies were calculated by solving the 2D Helmholtz equation [\refeq{eq:helmholtzScal}] for the circle as in Ref.~\cite{Bittner2009}. The difference between the measured and the calculated frequencies in \reffig{fig:freqFSRcomp}(a) is as large as half a FSR (i.e., the distance between two resonances with the same radial quantum number) and slowly decreases with increasing frequency. This is in accordance with the result that the FSR of the calculated resonances in \reffig{fig:freqFSRcomp}(b) is slightly larger than that of the measured ones. Since the frequency spectrum consists of series of almost equidistant resonances, the effect of this systematic error on the peak positions in the length spectrum can be estimated by considering a simple 1D system with equidistant resonances whose distance equals $\FSR$. The peak position in the corresponding length spectrum is $\lmax \propto 1 / \FSR$, and a deviation of $\Delta \FSR$ leads to an error,
\begin{equation} \frac{\Delta \lmax}{\lmax} = - \frac{\Delta \FSR}{\FSR} \, , \end{equation}
of the peak position. With $\Delta \FSR = \FSR_\mathrm{expt} - \FSR_\mathrm{calc} \approx -0.4$ MHz compared to $\FSR \approx 120$ MHz we expect a deviation of $3 \permil$ or $\Delta \lmax \approx 5$ mm in the peak positions, which agrees quite well with the magnitude of the deviations found in \reffig{fig:lspektExactHKreis}. For disk \dB, the comparison between the measured and the calculated FSR (not shown here) yields $\Delta \FSR / \FSR \approx -1.1 \permil$, which also agrees well with the deviations of $\Delta \lmax / \lmax \approx 1.5 \permil$ found in \reffig{fig:lspektExactBKreis}. Thus, we may conclude that the deviations between the peak positions of the experimental length spectrum and that of the trace formula indeed arise from the systematic error of the $\neff$ model. Unfortunately, we know of no general method to estimate the magnitude of this systematic error beforehand. It should be noted that the exact magnitude of the systematic error contributing to the deviations found in Figs.\ \ref{fig:lspektExactHKreis} and \ref{fig:lspektExactBKreis} depends on the index of refraction used in the calculations. Still, it was shown in Ref.~\cite{Bittner2009} and also checked here that deviations remain regardless of the value of $n$ used, which is known with per mill precision for the disks \dA~and \dB~\cite{Classen2010}. Furthermore, \reffig{fig:freqFSRcomp}(b) demonstrates that the index of refraction of a disk cannot be determined without systematic error from the measured $\FSR$ even if the dispersion of $\neff$ is fully taken into account.

\section{\label{sec:conc}Conclusions}
The resonance spectra of two circular dielectric microwave resonators were measured and the corresponding length spectra were investigated. In contrast to previous experiments with 2D resonators \cite{Bittner2010}, flat 3D resonators were used. The length spectra were compared to a combination of the semiclassical trace formula for 2D dielectric resonators proposed in Ref.~\cite{Bogomolny2008} and a 2D approximation of the Helmholtz equation for flat 3D resonators using an effective index of refraction (in accordance with Ref.\ \cite{Lebental2007}). The experimental length spectra and the trace formula showed good agreement, and the different contributions of the POs to the length spectra could be successfully identified. The positions of the peaks in the experimental length spectrum are, however, slightly shifted with respect to the geometrical lengths of the POs. We found that this shift is related to two different effects, which are, first, the frequency dependence of the effective index of refraction
and, second, a systematic inaccuracy of the $\neff$ approximation. In the examples considered here, the former effect is as large as $5 \permil$ of the PO length while the latter effect is as large as $2 \permil$ of $\lpo$, and the two effects cancel each other in part. The results and methods presented here provide a refinement of the techniques used in Refs.\ \cite{Lebental2007, Bogomolny2011} and allow for the detailed understanding of the spectra of realistic microcavities and -lasers in terms of the 2D trace formula. Furthermore, many of the effects discussed here also apply to 2D systems made of a dispersive material. Some open problems remain, though. The comparison of the semiclassical trace formula with the exact one for the circle showed that the former needs to be improved for POs close to the critical angle. Furthermore, there are some deviations between the experimental length spectra and the trace formula predictions due to the systematic error of the effective index of refraction approximation. Its effect on the length spectra proved to be rather small and, thus, allowed for the identification of the different PO contributions. However, the computation of the resonance frequencies of flat 3D resonators based on the combination of the 2D trace formula and the $\neff$ approximation would lead to the same systematic deviations from the measured ones as in Ref.~\cite{Bittner2009}. Another problem with this systematic error is that its magnitude cannot be estimated \textit{a priori}. The comparison of the results for disk \dA~and disk \dB~seems to indicate that it gets smaller for $b / R \rightarrow 0$, but there are not enough data to draw final conclusions yet, especially since the value of $b / R$ is of similar magnitude for both disks. In fact, Ref.~\cite{Bittner2009} rather indicates that the systematic error of the $\neff$ model increases with decreasing $b / R$. This could be attributed to diffraction effects at the boundary of the disks that become more important when $b$ gets smaller compared to the wavelength. On the other hand, the exact 2D case is recovered for $b \rightarrow \infty$, i.e., an infinitely long cylinder. In conclusion the accuracy of the $\neff$ model in the limit $b / R \rightarrow 0$ remains an open problem. An analytical approach to the problem of flat dielectric cavities that is more accurate than the $\neff$ model would be of great interest. Another perspective direction is to consider 3D dielectric cavities with the size of all sides having the same order of magnitude and to develop a trace formula for them similar to those for metallic 3D cavities \cite{Balian1977, Frank1996, Dembowski2002}.

\begin{acknowledgments}
The authors wish to thank C. Classen from the department of electrical engineering and computer science of the TU Berlin for providing numerical calculations to validate the measured resonance data. This work was supported by the DFG within the Sonderforschungsbereich 634.
\end{acknowledgments}

\appendix

\section{\label{sec:polMeasCalc}Influence of the metallic plate on the resonance frequencies}

\begin{figure}[!b]
\begin{center}
\includegraphics[width = 6.4 cm]{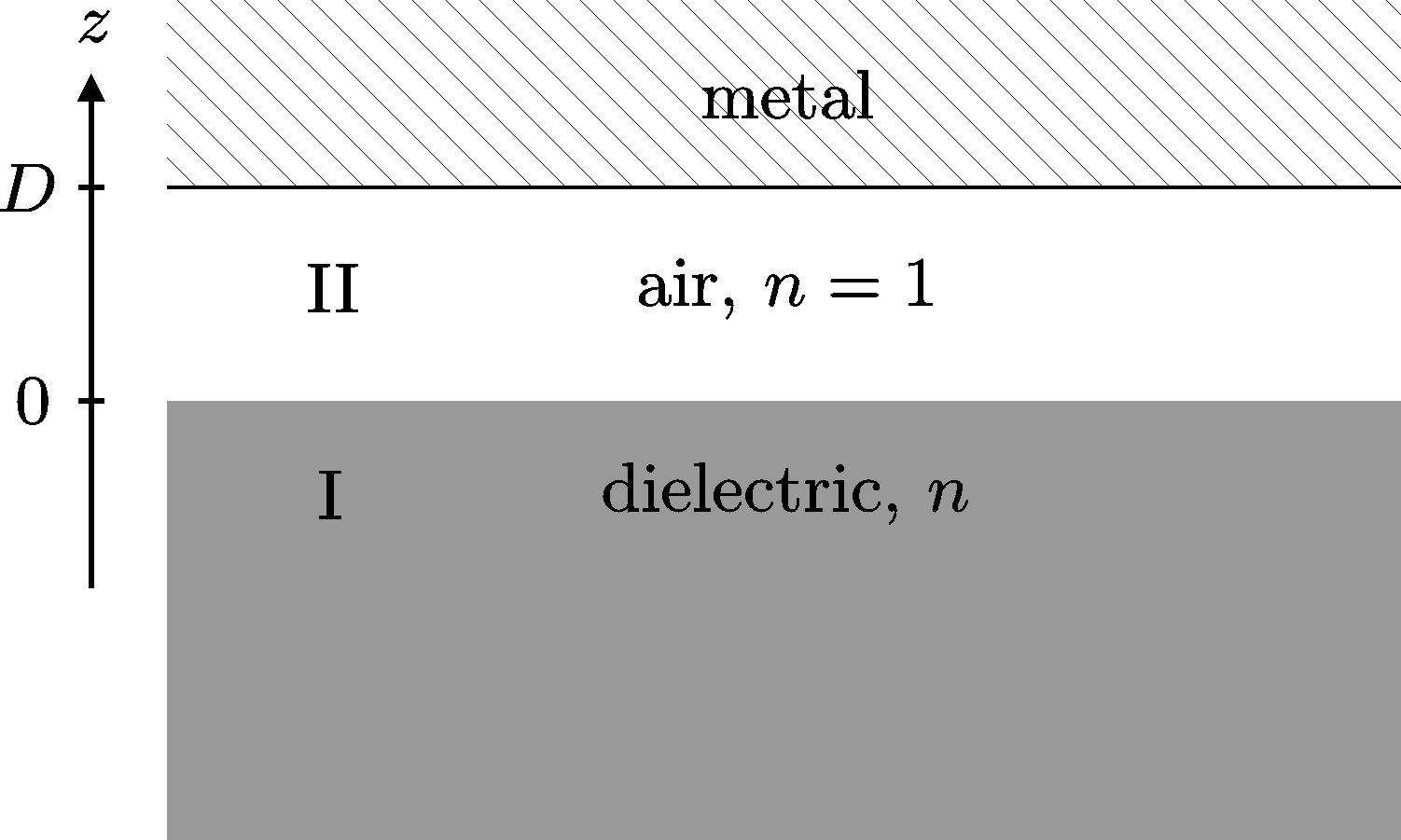}
\end{center}
\caption{\label{fig:fresnelSetup}Geometry and coordinate system for the calculation of the reflection coefficient $r(\dist)$.}
\end{figure}

The placement of a metal plate parallel to the dielectric disk influences the effective index of refraction, thus, leading to a shift of the resonance frequencies. In order to determine the change of $\neff$, we, first, calculate the reflection coefficient $r(D)$ for a wave traveling inside the dielectric and being reflected at the dielectric-air interface with the metal plate at a distance $\dist$. The geometry used here is depicted in \reffig{fig:fresnelSetup}. The ansatz for the $E_z$ field (TM polarization) and, respectively, the $B_z$ field (TE polarization) is
\begin{equation} \begin{array}{lcl} \mathrm{A} \Psi(x, y)(e^{i k_z z} + r e^{-i k_z z}) e^{-i \omega t} & : & z \leq 0 \\ \Psi(x, y) (b_1 e^{-q z} + b_2 e^{+q z}) e^{-i \omega t} & : & 0 \leq z \leq \dist \, , \end{array} \end{equation}
where $\Psi$ fulfills $(\Delta + \kt^2) \Psi = 0$, $\omega = 2 \pi f$ is the angular frequency, $\mathrm{A}$, $b_{1, 2}$ are constants, and $r$ is the reflection coefficient. The different wave numbers are connected by
\begin{equation} k^2 = (\kt^2 + k_z^2) / n^2 = \kt^2 - q^2 = \kt^2 / \neff^2 \, . \end{equation}
For the case of TIR considered here, $q$ is real, and the penetration depths $\depth$ of the field intensity into region II is $\depth = 1 / (2 q)$. For the case of TM polarization the electric field is
\begin{equation} E_z^\mathrm{(II)} = \mathrm{B} \Psi \cosh{[q(z - \dist)]} e^{-i \omega t} \end{equation}
since it obeys Neumann boundary conditions at the metal plate, where $\mathrm{B}$ is a constant. The boundary conditions at the interface between region I and II are that $n^2 E_z$ and $\pbyp{E_z}{z}$ are continuous, which yields
\begin{equation} r_\mathrm{TM}(\dist) = \frac{k_z - i n^2 q \tanh{(q D)}}{k_z + i n^2 q \tanh{(q D)}} \, . \end{equation}
For the case of TE polarization the magnetic field in region II is
\begin{equation} B_z^\mathrm{(II)} = \mathrm{B} \Psi \sinh{[q (z - \dist)]} e^{-i \omega t} \end{equation}
since it obeys Dirichlet boundary conditions at the metal plate. With the condition that $B_z$ and $\pbyp{B_z}{z}$ are continuous at the dielectric interface,
\begin{equation} r_\mathrm{TE}(\dist) = \frac{k_z - i q \coth{(q \dist)}}{k_z + i q \coth{(q \dist)}} \end{equation}
is obtained. Analogous to \refeq{eq:fresnelPhase} we define $r(\dist) = \exp{[-2 i \delta(\dist)]}$ and obtain
\begin{equation} \tan{[\delta(\dist)]} = \nu \sqrt{\frac{\neff^2 - 1}{n^2 - \neff^2}} h[\dist / (2 \depth)] \end{equation}
with 
\begin{equation} h(x) = \left\{ \begin{array}{lcl} \tanh{x} & : & \mathrm{TM \,\, polarization} \\ \coth{x} & : & \mathrm{TE \,\, polarization} \, . \end{array} \right. \end{equation}
For $\dist \gg \depth$, \refeq{eq:fresnelPhase} is recovered. This explains why the optical table does not disturb the resonator. By inserting $r(\dist)$ into \refeq{eq:kzQbed} we obtain
\begin{equation} k b = \frac{1}{\sqrt{n^2 - \neff^2}} \{ \delta_0 + \delta(\dist) + \zeta \pi \} = g(\neff, \dist) \end{equation}
as condition for $\neff(\dist)$, where $\delta_0$ refers to \refeq{eq:fresnelPhase}. In order to determine the effect of a change of $\dist$ on the effective index of refraction, we compute
\begin{equation} \pbyp{\neff}{\dist} = \left. - \frac{1}{\sqrt{n^2 - \neff^2}} \pbyp{\delta(\dist)}{\dist} \middle/ \pbyp{g}{\neff} \right. \, , \end{equation}
where
\begin{equation} \pbyp{\delta(\dist)}{\dist} = \frac{\tan{\delta_0}}{2 \depth [1 + h^2 \tan^2{(\delta_0)}]} h'[\dist/(2 \depth)] \, . \end{equation}
The derivative of $h$ is
\begin{equation} h'(x) = \left\{ \begin{array}{rcl} \cosh^{-2}{(x)} & : & \mathrm{TM} \\ -\sinh^{-2}{(x)} & : & \mathrm{TE} \end{array} \right. , \end{equation}
which approaches $h'(x) = 4 v e^{-2x}$ for $x \rightarrow \infty$ with $v = +1$ for TM and $v = -1$ for TE polarization. Then, for $\dist \gg \depth$,
\begin{equation} \pbyp{\neff}{\dist} = - \frac{C}{\depth} v e^{-\dist / \depth} \end{equation}
with $C > 0$ given as
\begin{equation} C = \frac{1}{\pbyp{g}{\neff}} \frac{2}{\sqrt{n^2 - \neff^2}} \frac{\tan{\delta_0}}{1 + h^2 \tan^2{(\delta_0)}} \, . \end{equation}
Accordingly, for large distances of the metal plate,
\begin{equation} \label{eq:neffOfD} \neff(\dist) \approx \left. \neff \right|_{\dist = \infty} + \left\{ \begin{array}{lcl} +C e^{-\dist / \depth} & : & \mathrm{TM} \\ -C e^{-\dist / \depth} & : & \mathrm{TE} \, , \end{array} \right. \end{equation}
i.e., for TM modes $\neff$ increases for decreasing distance $\dist$ of the metal plate from the resonator and for TE modes it gets smaller. This qualitative behavior is found for the whole range of $\dist$, even when \refeq{eq:neffOfD} is no longer valid. Since the resonance frequencies of the disk are to first-order approximation $\propto 1 / \neff$, the TM modes are shifted to lower and the TE modes to higher frequencies as observed in \reffig{fig:polMeas}. \newline

\section{\label{sec:exactTraceForm}Exact trace formula for the circular dielectric resonator}
The DOS for the TM modes of the 2D circular dielectric resonator is
\begin{equation} \rho(k) = - \frac{1}{\pi} \sum \limits_{m = -\infty}^{+ \infty} \sum \limits_{n_r = 1}^{\infty} \Im{ \frac{1}{k - k_{m, \, n_r}} } \end{equation}
where the eigenvalues $k_{m, \, n_r}$ are the roots of \cite{Hentschel2002b}
\begin{equation} s_m(x) = x [ n \besseljp{m}{nx} \hankel{m}{1}{x} - \besselj{m}{nx} \hankelp{m}{1}{x} ] \end{equation}
with $x = k R$. The factor $x$ ensures that $s_m$ has no poles. Informally one can write $s_m(x) = f(x) \Pi_{n_r} (x - R k_{m, n_r})$, where $f(x)$ is a certain smooth function without zeros and poles. This means that the DOS can be written as
\begin{equation} \label{eq:rhoViaSm} \rho(k) = - \frac{R}{\pi} \sum \limits_{m = -\infty}^{+ \infty} \Im{ \frac{\dbyd{s_m}{x}}{s_m} } + \, \mathrm{smooth\,term.} \end{equation}
The smooth term contributes to the Weyl expansion and requires a separate treatment \cite{Bogomolny2008, Bogomolny2011}. In the following we ignore all such terms. The derivative $\dbyd{s_m}{x}$ in \refeq{eq:rhoViaSm} contains two terms,
\begin{equation} \dbyd{s_m}{x} = \pbyp{s_m}{x} + \pbyp{s_m}{n} \dbyd{n}{x} \, . \end{equation}
The first term is
\begin{equation} \pbyp{s_m}{x} = -x (n^2 - 1) \besselj{m}{nx} \hankel{m}{1}{x} \end{equation}
where the second derivatives of the Bessel and Hankel functions were resolved via the Bessel differential equation. The second term is
\begin{equation} \begin{array}{rcl} \pbyp{s_m}{n} & = & n x^2 \besselj{m}{nx} \hankel{m}{1}{x} \times \\ & & \left[ \frac{m^2}{n^2 x^2} - 1 - \frac{1}{n^2} \left( \frac{\hankelsp{m}{1}}{\hankels{m}{1}} \right)^2(x) \right] \end{array} \end{equation}
where $\frac{\besseljsp{m}}{\besseljs{m}}(nx)$ was replaced by $\frac{1}{n} \frac{\hankelsp{m}{1}}{\hankels{m}{1}}(x)$ since $s_m(x) = 0$. We approximate
\begin{equation} \frac{\hankelsp{m}{1}}{\hankels{m}{1}}(x) \approx - \sqrt{\frac{m^2}{x^2} -1} \end{equation}
and obtain
\begin{equation} \pbyp{s_m}{n} = - \frac{x^2}{n} (n^2 - 1) \besselj{m}{nx} \hankel{m}{1}{x} \, . \end{equation}
It was checked numerically that this approximation is very precise, which is why we still call the result an exact trace formula. Combining both terms yields
\begin{equation} \frac{\dbyd{s_m}{x}}{s_m} = - \frac{(n^2 - 1)}{\tilde{s}_m(x)} \left(1 + \frac{k}{n} \dbyd{n}{k} \right) \end{equation}
with
\begin{equation} \tilde{s}_m(x) = n \frac{\besseljsp{m}}{\besseljs{m}}(nx) - \frac{\hankelsp{m}{1}}{\hankels{m}{1}}(x) \, . \end{equation}
The DOS is, thus,
\begin{equation} \rho(k) = \frac{R (n^2 - 1) \left( 1 + \frac{k}{n} \dbyd{n}{k} \right)}{2 \pi i}  \sum \limits_{m = - \infty}^{+ \infty} \frac{1}{\tilde{s}_m(x)} + \mathrm{c.c.} \, . \end{equation}
We replace the Bessel functions $\besseljs{m}$ in $\tilde{s}_m$ by $(\hankels{m}{1} + \hankels{m}{2}) / 2$ and extract a term $[E_m(x) + 1]$ with $E_m$ defined in \refeq{eq:treEm} to obtain
\begin{equation} \tilde{s}_m = \frac{B_m}{E_m + 1} [1 - E_m R_m] \, , \end{equation}
where $R_m$, $A_m$, and $B_m$ are defined in Eqs.~(\ref{eq:treRm}), (\ref{eq:treAm}), and (\ref{eq:treBm}), respectively. With the help of the geometric series $1 / \tilde{s}_m$ is rewritten as
\begin{equation} \frac{1}{\tilde{s}_m} = \frac{1}{B_m} + \tilde{P}_m \sum \limits_{q = 1}^{\infty} (E_m R_m)^q \end{equation}
with
\begin{equation} \tilde{P}_m = \left( 1 + \frac{1}{R_m} \right) B_m \, . \end{equation}
Using the Wronskian \cite{Bateman1953}
\begin{equation} W[\hankel{m}{2}{z}, \hankel{m}{1}{z}] = \frac{4 i}{\pi z} \end{equation}
this simplifies to
\begin{equation} \tilde{P}_m = \frac{4 i}{\pi x A_m B_m \hankel{m}{1}{nx} \hankel{m}{2}{nx}} \end{equation}
and we obtain
\begin{equation} \begin{array}{rcl} \rho(k) & = & \frac{R (n^2 - 1)}{2 \pi i} \left( 1 + \frac{k}{n} \dbyd{n}{k} \right) \\ \\ & & \times \sum \limits_{m = -\infty}^{+ \infty} \left[ \frac{1}{B_m} + \tilde{P}_m \sum \limits_{q = 1}^{\infty} (R_m E_m)^q \right] + \mathrm{c.c.} \, . \end{array} \end{equation}
The first term, $1 / B_m$, corresponds to the smooth part of the DOS. Since we are only interested in the fluctuating part, we drop the term and apply the Poisson resummation formula to the rest to obtain
\begin{equation} \rhof(k) = \frac{2 R}{\pi^2 x} \sum \limits_{\eta = -\infty}^{+ \infty} \int \limits_{-\infty}^{+\infty} dm \, e^{2 \pi i m \eta} P_m \sum \limits_{q = 1}^\infty (E_m R_m)^q + \mathrm{c.c.} \end{equation}
with $P_m$ defined in \refeq{eq:trePm}. Replacing $\sum \limits_{\eta = -\infty}^{+ \infty} \sum \limits_{q = 1}^\infty$ with $2 \sum \limits_\po$ and ignoring those $(q, \eta)$ combinations that are not related to any POs and, thus, do not give significant contributions finally yields \refeq{eq:trFormExact}. Taking the semiclassical limit as described in Ref.\ \cite{Bogomolny2008} results in \refeq{eq:trFormScl} with an additional factor of $(1 + \frac{k}{n} \dbyd{n}{k})$. This means that the dispersion of $n$ leads to slightly higher amplitudes in the semiclassical limit.

\end{document}